\documentclass[journal]{IEEEtran}
\usepackage{bm}
\usepackage{cite}
\usepackage{amsmath,amssymb,amsfonts}
\usepackage{algorithmic}
\usepackage{graphicx}
\usepackage{textcomp}
\usepackage{color}
\usepackage{subfigure}

\def\BibTeX{\rm B\kern-.05em{\sc i\kern-.025em b}\kern-.08emT\kern-.1667em\lower.7ex\hbox{E}\kern-.125emX}
\allowdisplaybreaks[4]
\newtheorem{assumption}{Assumption}
\newtheorem{lemma}{Lemma}
\newtheorem{proposition}{Proposition}
\newtheorem{theorem}{Theorem}
\begin{document}

\title{IRS Assisted Federated Learning: A Broadband Over-the-Air Aggregation Approach}

\author{Deyou Zhang,~\IEEEmembership{Member,~IEEE}, Ming Xiao,~\IEEEmembership{Senior Member,~IEEE}, Zhibo Pang,~\IEEEmembership{Senior Member,~IEEE}, \\Lihui Wang, and H. Vincent Poor,~\IEEEmembership{Life Fellow,~IEEE} \vspace{-2em}

\thanks{D. Zhang and M. Xiao are with the Division of Information Science and Engineering (ISE), Royal Institute of Technology (KTH), Stockholm 10044, Sweden (email: \{deyou, mingx\}@kth.se). Z. Pang is with ABB Corporate Research, V\"{a}ster{\aa}s 72226, Sweden, and also with the Division of ISE, KTH, Stockholm 10044, Sweden (email: pang.zhibo@se.abb.com, zhibo@kth.se). L. Wang is with the Department of Production Engineering, KTH, Stockholm 10044, Sweden (email: lihui.wang@iip.kth.se). H. V. Poor is with the Department of Electrical and Computer Engineering, Princeton University, Princeton, NJ 08544, USA (poor@princeton.edu).}
}

\maketitle

\begin{abstract}
We consider a broadband over-the-air computation empowered model aggregation approach for wireless federated learning (FL) systems and propose to leverage an intelligent reflecting surface (IRS) to combat wireless fading and noise. We first investigate the conventional node-selection based framework, where a few edge nodes are dropped in model aggregation to control the aggregation error. We analyze the performance of this node-selection based framework and derive an upper bound on its performance loss, which is shown to be related to the selected edge nodes. Then, we seek to minimize the mean-squared error (MSE) between the desired global gradient parameters and the actually received ones by optimizing the selected edge nodes, their transmit equalization coefficients, the IRS phase shifts, and the receive factors of the cloud server. By resorting to the matrix lifting technique and difference-of-convex programming, we successfully transform the formulated optimization problem into a convex one and solve it using off-the-shelf solvers. To improve learning performance, we further propose a weight-selection based FL framework. In such a framework, we assign each edge node a proper weight coefficient in model aggregation instead of discarding any of them to reduce the aggregation error, i.e., amplitude alignment of the received local gradient parameters from different edge nodes is not required. We also analyze the performance of this weight-selection based framework and derive an upper bound on its performance loss, followed by minimizing the MSE via optimizing the weight coefficients of the edge nodes, their transmit equalization coefficients, the IRS phase shifts, and the receive factors of the cloud server. Furthermore, we use the MNIST dataset for simulations to evaluate the performance of both node-selection and weight-selection based FL frameworks.
\end{abstract}


\IEEEpeerreviewmaketitle

\section{Introduction}
\IEEEPARstart{R}{ecent} advances in artificial intelligence (AI), edge computing, and 5G networking have promoted the rapid proliferation of the Industrial Internet of Things (IIoT) \cite{IIoT-Magazine, IIoT-Tutorial}. In IIoT systems, each front-end device (e.g., sensor or camera) continuously generates a large amount of data, which often needs to be processed with AI or machine learning methods \cite{IIoT-Magazine}. Taking the automatic sorting system as an example, the images of industrial products on assembly lines are firstly captured by multiple cameras from different angles and then uploaded to a remote server for processing and analysis. Due to the enormous data volumes and the limited transmission capability of consumer-grade cameras, such a data offloading process is time-demanding \cite{FL-IIoT-Magazine}. More importantly, the data from many confidential products often involve sensitive information. For example, the image of a product/object can reflect a lot of information about this product/object such as shape, color, size, etc. Thus, collecting IIoT data to a centralized server can also lead to privacy problems \cite{Blockchain-Qu, Blockchain-ZhangYan}.

\begin{table*}[!htbp]
\centering
\caption{A comparison between our and existing works.} \label{Comparison-Table}
\begin{tabular}{|c|c|c|c|c|}
\hline
\textbf{Literature} & \cite{HangLiu-TWC} & \cite{ZhibinWang-TWC} & \cite{YuanweiLiu-IoTJ} & Our Work \\
\hline
\textbf{Server Configuration}  & Multi-Antenna & Multi-Antenna & Single-Antenna & Single-Antenna \\
\hline
\textbf{IRS Phase Shift Design} & SCA & Matrix Lifting + SCA & SCA & Matrix Lifting + SCA \\
\hline
\textbf{Edge Node Selection} & Gibbs Sampling & Bisection Search & SCA & SCA \\
\hline
\textbf{Optimization Manner} & Joint & Two-Step & Alternative & Joint \\
\hline
\textbf{System Configuration} & Narrowband & Narrowband & Narrowband & Wideband \\
\hline
\textbf{Proposed New FL Paradigm} & No & No & No & Yes \\
\hline
\end{tabular}
\vspace{-1em}
\end{table*}

To cope with the dual challenges of big data and privacy protection, federated learning (FL) provides a new paradigm for time-sensitive and privacy-preserving IIoT applications \cite{FL-IIoT-Magazine}. A typical FL-IIoT framework consisting of multiple front-end devices and a cloud server executes the following two procedures until convergence is achieved. 1) \emph{model broadcast}: the cloud server broadcasts a global model to the front-end devices, each of which computes a local gradient vector leveraging its private dataset; 2) \emph{model aggregation}: the front-end devices upload their computed local gradient vectors to the cloud server for aggregation. Since only model parameters rather than the raw data are uploaded to the cloud server, FL is capable of reducing communication costs and achieving privacy protection \cite{FL-IIoT-Magazine}.

Despite the advantages of FL, the uplink model aggregation procedure is a severe bottleneck for FL training in communication overhead, particularly in wireless scenarios \cite{WFL-CL}. To alleviate this problem, several works have proposed to optimize resource allocation among the front-end devices to enhance FL learning efficiency \cite{RaFL, HaoChen-IoTJ, MingzheChen-TWC}. However, those literatures \cite{RaFL, HaoChen-IoTJ, MingzheChen-TWC} considered orthogonal multiple access (OMA) protocols for model aggregation, such that the required wireless resources, e.g., bandwidth, increased linearly with the number of devices involved in FL. As a result, when many front-end devices are involved in FL, the model aggregation procedure would incur excessive resource consumption \cite{WFL-CL2}.

To improve the communication efficiency in model aggregation, over-the-air computation (AirComp) empowered model aggregation approach has emerged \cite{GuangxuZhu-TWC, Deniz-TWC, Deniz-TSP, YuanmingShi-TWC, ShuaiWang-TWC, YongmingHuang-JSAC}. In such an approach, the devices use the same time-frequency resources to upload their local gradient vectors (or model updates) to the cloud server, which implements model aggregation by exploiting the waveform superposition property of multiple-access channels. Specifically, the first AirComp-empowered model aggregation research appeared in \cite{GuangxuZhu-TWC}, where the authors derived two tradeoffs between communication and learning metrics and demonstrated that AirComp indeed substantially reduces the model uploading latency compared to the OMA protocols. To further reduce the uploading overhead, the authors in \cite{Deniz-TWC} and \cite{Deniz-TSP} proposed to first ``sparsify'' and compress the local gradient vectors and then upload them to the cloud server for model aggregation.

Although AirComp is envisioned as a scalable model aggregation paradigm, it still suffers from the ``straggler'' problem, i.e., the devices with poor channel conditions dominate the model aggregation error. To alleviate this problem, literatures \cite{YuanmingShi-Network, ZhibinWang-TWC, HangLiu-TWC, YuanweiLiu-IoTJ, YuanweiLiu-TWC} have proposed to employ intelligent reflecting surfaces (IRSs) to enhance the channels between frond-end devices and the cloud server. As reported in \cite{Renzo-Access, RuiZhang-Magazine, ChongwenHuang-TWC}, the IRS is a cost-effective technology to overcome the detrimental effect of channel fading in wireless communications. Precisely, an IRS consists of a row of passive reflecting elements, and by adjusting the phase shifts of these elements, we can control the propagation of the reflected signal, making it superpose constructively with the signal over the direct link to strengthen the received signal power \cite{RuiZhang-Magazine}. Since the purpose of FL is different from that of traditional communication systems, the conventional transmit/receive/reflect designs for IRS-assisted communication systems cannot apply to the IRS-assisted FL systems, whose transmit/receive/reflect strategies need to be redesigned. The recent works in \cite{YuanmingShi-Network, ZhibinWang-TWC, HangLiu-TWC, YuanweiLiu-IoTJ, YuanweiLiu-TWC} show that IRSs indeed can alleviate the detrimental effect of channel fading and communication noise on model aggregation. Compared to FL systems without IRSs, considerable performance improvements were observed in these works \cite{YuanmingShi-Network, ZhibinWang-TWC, HangLiu-TWC, YuanweiLiu-IoTJ, YuanweiLiu-TWC}.

The ``multiplicative fading'' effect limits the benefit of passive IRSs \cite{LinglongDai-ActiveRIS}, and the straggler problem still exists in IRS-assisted FL systems \cite{YuanmingShi-Network, ZhibinWang-TWC, HangLiu-TWC, YuanweiLiu-IoTJ, YuanweiLiu-TWC}. As such, those state-of-the-art works also proposed to discard stragglers from model aggregation to avoid severe aggregation errors. In particular, the authors in \cite{ZhibinWang-TWC, HangLiu-TWC, YuanweiLiu-IoTJ} investigated a joint device selection and transmit/receive/passive beamforming design to enhance the FL learning performance. However, discarding devices reduces the total number of training data samples, which inevitably compromises the performance of FL, particularly when the discarded devices possess unique features.

To avoid this dilemma, instead of discarding any front-end devices (termed edge nodes below), we assign each of them a carefully designed weight coefficient in model aggregation to control the aggregation error. In other words, amplitude alignment of the received local gradient parameters from different edge nodes is not required, which is different from those state-of-the-art works \cite{ZhibinWang-TWC, HangLiu-TWC, YuanweiLiu-IoTJ}. Moreover, since future communication systems are wideband, we thus propose to implement model aggregation over broadband channels, which is also different from \cite{ZhibinWang-TWC, HangLiu-TWC, YuanweiLiu-IoTJ}. A detailed comparison between our work and \cite{ZhibinWang-TWC, HangLiu-TWC, YuanweiLiu-IoTJ} is provided in Table \ref{Comparison-Table}. The contributions of this paper are summarized as follows.

1) Focusing on IRS-assisted FL systems, we consider the broadband AirComp-empowered model aggregation approach. We first study the conventional node-selection based FL framework, where a few edge nodes are dropped in model aggregation to control the aggregation error. We theoretically analyze the convergence performance of this node-selection based framework and derive an upper bound on its performance loss, i.e., the expected difference between the training loss and the optimal loss, which is shown to be related to the selected edge nodes.

2) Subsequently, we minimize the mean-squared error (MSE) between the desired aggregated gradient vector and the actually received one by optimizing edge node selection, transceiver design, and IRS configuration, which is a highly intractable combinatorial optimization problem. By using the matrix lifting technique and difference-of-convex (DC) programming, we successfully transform the original intractable optimization problem into a convex one and solve it using off-the-shelf solvers.

3) To avoid a noticeable decrease in the learning performance caused by node selection, we further propose a weight-selection based FL framework. In such a framework, we assign each edge node a carefully designed weight coefficient in model aggregation instead of discarding any of them. As in the node-selection based framework, we analyze the performance of this weight-selection based framework and derive an upper bound on its performance loss. We also minimize the MSE by jointly optimizing the weight coefficients of the edge nodes, their transmit equalization coefficients, the IRS phase shifts, and the receive factors of the cloud server.

4) We use the MNIST dataset for simulations to evaluate the performance of both node-selection and weight-selection based FL frameworks. Simulation results show that the IRS indeed alleviates the straggler problem, and the weight-selection based framework achieves higher prediction accuracy than its node-selection based counterpart.

\begin{table}[!t]
\centering
\caption{Summary of major notations.}\label{Notation-Table}
$~$\begin{tabular}{|c|l|}
\hline
Notation            & Description \\
\hline
$\nabla$            & Gradient operator \\
\hline
$\mathbb E$         & Expectation operator \\
\hline
$\hat{\bm a}$       & Noisy version of $\bm a$ \\
\hline
$\|\cdot\|$         & $\ell_2$-norm \\
\hline
$\text{mat}$        & Reshape a vector into a matrix  \\
\hline
$\text{vec}$        & Reshape a matrix into a vector \\
\hline
$[\bm A]_{i,j}$     & $(i,j)$-th element of matrix $\bm A$ \\
\hline
$\mathcal {CN}(\bm{\mu}, \bm{\Sigma})$  & Complex Gaussian distribution with mean $\bm{\mu}$ and \\
                                        & covariance matrix $\bm{\Sigma}$ \\
\hline
$\frac{1}{\textsf{SNR}}$ & Noise variance \\
\hline
$\eta$              & Learning rate \\
\hline
$\bm w$             & Global model parameter vector \\
\hline
$\bm g / \bm G$     & Global gradient vector/matrix \\
\hline
$d$                 & Dimension of the model parameter vector \\
\hline
$N$                 & Total number of sub-channels \\
\hline
$T$                 & Total number of time slots \\
\hline
$\cal K$            & Set of the $K$ edge nodes \\
\hline
${\cal S}^{[l]}$    & Selected subset of edge nodes at the $l$-th round \\
\hline
$\bm g_k / \bm G_k$ & Local gradient vector/matrix of edge node $k$ \\
\hline
$\bar g_k$          & Mean value of the elements in $\bm g_k$ \\
\hline
$\delta_k$          & Variance of the elements in $\bm g_k$ \\
\hline
$q_k$               & Weight coefficient of edge node $k$ \\
\hline
${\cal D}_k$        &Local data set of edge node $k$ \\
\hline
$m_j$               & Receive factor at the $j$-th sub-channel \\
\hline
$b_{k, j}$          & Transmit equalization coefficient of edge node \\
                    & at the $j$-th sub-channel \\
\hline
$h_{e, k, j}$       & Equivalent channel between edge node $k$ and \\
                    & the cloud server \\
\hline
$F(\bm w)$          & Global loss function \\
\hline
$F_k(\bm w)$        & Local loss function at edge node $k$ \\
\hline
$\{\bm u_{k,i}, v_{k,i}\}$ & $i$-th sample in ${\cal D}_k$ \\
\hline
$f\left(\bm w; \bm u_{k,i}, v_{k,i}\right)$ & Sample-wise loss function \\
\hline
\end{tabular}
\vspace{-1em}
\end{table}

The remainder of this paper is organized as follows. Section \ref{Section-SystemModel} introduces preliminaries about FL, the node-selection based FL framework, the IRS-assisted communication model, and the AirComp-empowered model aggregation. Section \ref{Section-NSFL-Convergence} analyzes the convergence of the node-selection based FL framework. Section \ref{Section-NSFL-PF} describes how to optimize the selected edge nodes, their transmit equalization coefficients, the IRS phase shifts, and the receive factors of the cloud server to minimize the model aggregation error of the node-selection based FL framework. We develop the weight-selection based FL framework in Section \ref{Section-WSFL}. Simulation results are provided in Section \ref{Section-Results}, and this paper concludes in Section \ref{Section-Conclusions}. The major notations used in the article are listed in Table \ref{Notation-Table}.

\section{System Model} \label{Section-SystemModel}
In this section, we develop an AirComp-empowered model aggregation approach for FL, and an IRS is employed to enhance the wireless transmissions.

\subsection{Preliminaries of FL}
\begin{figure*}
\centering
\subfigure[Model broadcast]
{\includegraphics[width = 5.6cm]{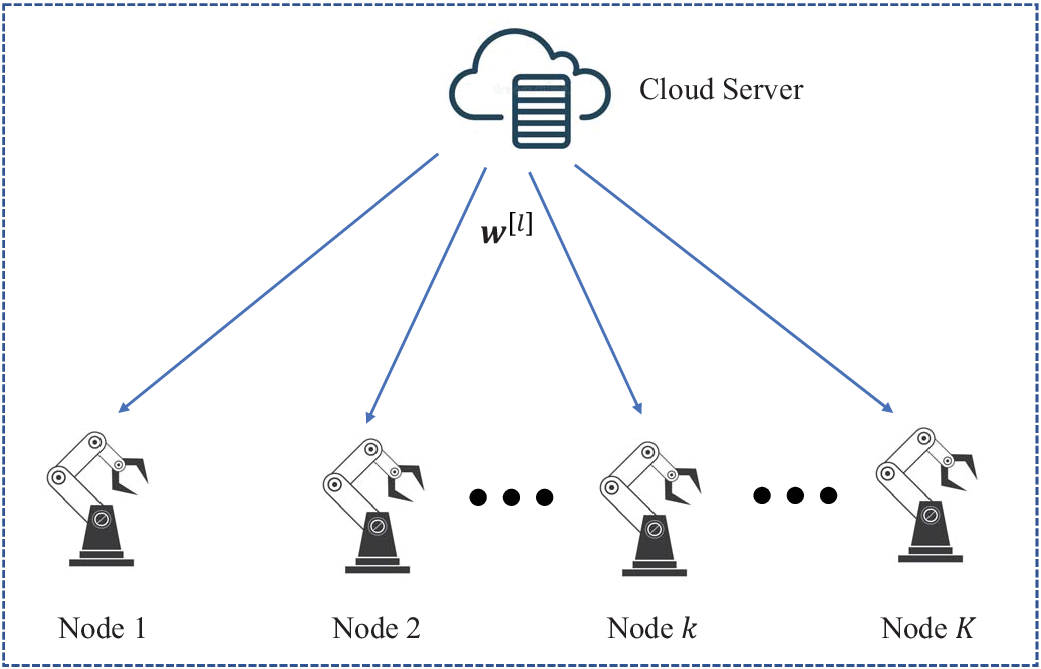}\label{MB}}
\subfigure[Node-selection based model aggregation]
{\includegraphics[width = 5.6cm]{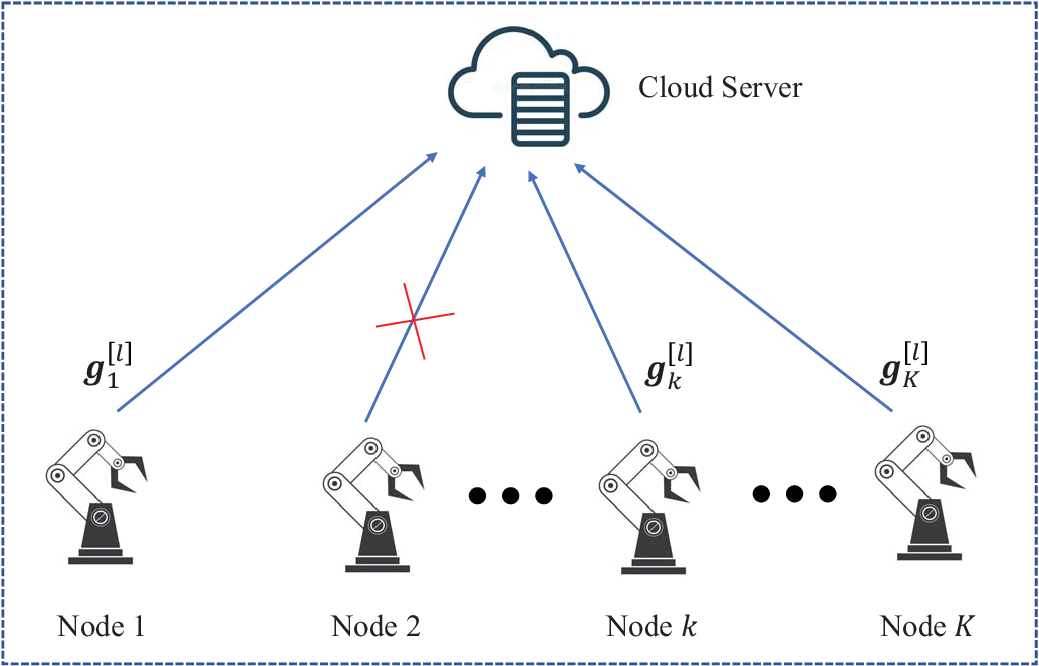}\label{CMA}}
\subfigure[Weight-selection based model aggregation]
{\includegraphics[width = 5.6cm]{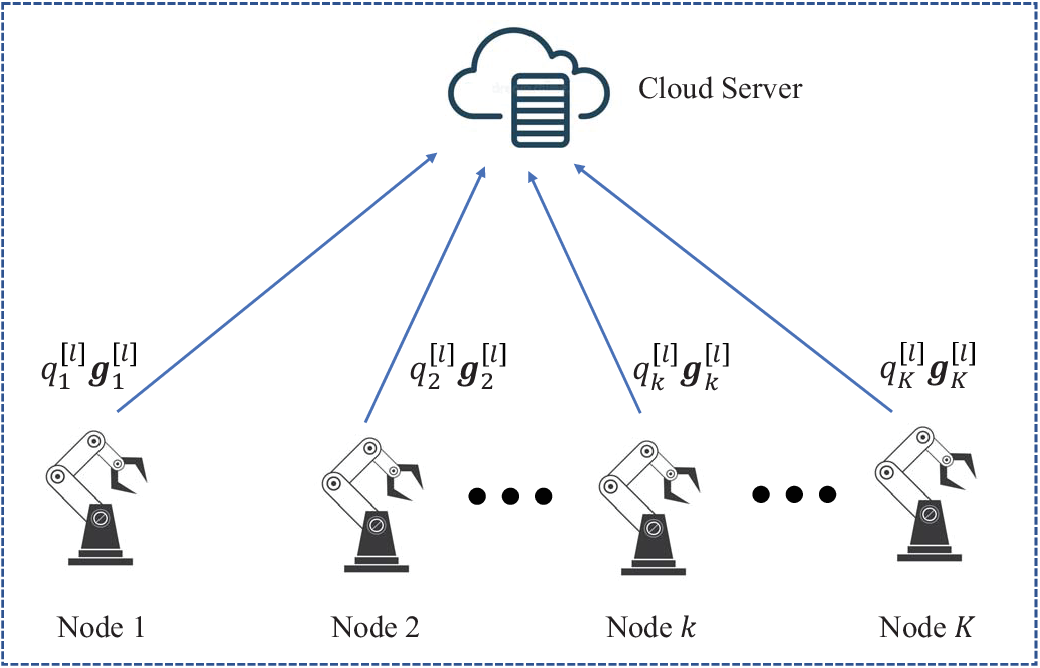}\label{PMA}}
\caption{Illustration of the $l$-th FL training round.}\label{FL}
\end{figure*}

As illustrated in Fig. \ref{FL}, a typical FL system consists of a cloud server and $K$ edge nodes. Edge node $k$ has a local dataset ${\cal D}_k$ that contains $D_k = |{\cal D}_k|$ labeled data samples, denoted by $\{\bm u_{k,i}, v_{k,i}\}$, $\forall i = 1, \cdots, D_k$, and $\forall k \in {\cal K} \triangleq \{1, \cdots, K\}$. The tuple $\{\bm u_{k,i}, v_{k,i}\}$ represents the $i$-th data sample in ${\cal D}_k$, consisting of feature vector $\bm u_{k,i}$ and its ground-truth label $v_{k,i}$. The learning objective of FL is to seek a model parameter vector $\bm w \in {\cal R}^d$ that can minimize the following global loss function
\begin{equation}\label{Eq-GLF}
    F\left(\bm w\right) = \frac{1}{\sum\nolimits_{j=1}^K D_j} \sum\limits_{k=1}^K \sum\limits_{i=1}^{D_k} f\left(\bm w; \bm u_{k,i}, v_{k,i}\right),
\end{equation}
in a distributed manner, where $f\left(\bm w; \bm u_{k,i}, v_{k,i}\right)$ is termed sample-wise loss function quantifying the misfit of $\bm w$ on the data sample $\{\bm u_{k,i}, v_{k,i}\}$ \cite{MPFriedlander}.

To this end, we follow \cite{RaFL, GuangxuZhu-TWC, ZhibinWang-TWC, HangLiu-TWC}, and define the local loss function of $\bm w$ on ${\cal D}_k$ as
\begin{equation}\label{LLF}
    F_k\left(\bm w\right) = \frac{1}{D_k} \sum\limits_{i=1}^{D_k} f\left(\bm w; \bm u_{k,i}, v_{k,i}\right).
\end{equation}
Then, the global loss function in \eqref{Eq-GLF} can be rewritten as
\begin{equation}\label{Eq-GLF2}
    F\left(\bm w\right) = \frac{1}{\sum\nolimits_{j=1}^K D_j} \sum\limits_{k=1}^K D_k F_k\left(\bm w\right).
\end{equation}
Following \cite{GuangxuZhu-TWC, ZhibinWang-TWC}, we further assume that the $K$ local datasets have equal size, i.e., $D_k = D$, $\forall k \in {\cal K}$, such that $F\left(\bm w\right)$ in \eqref{Eq-GLF2} reduces to
\begin{equation}
    F\left(\bm w\right) = \frac{1}{K} \sum\limits_{k=1}^K F_k\left(\bm w\right).
\end{equation}

\subsection{Node-Selection Based FL Systems}
In FL systems, the model parameter vector $\bm w$ is computed in an iterative manner between the cloud server and the $K$ edge nodes, which is repeated for a fixed number of $L$ rounds or until a global consensus is achieved. Precisely, the $l$-th round of a conventional FL system consists of the following procedure.

\textbf{Model broadcast}: The cloud server broadcasts the current global model parameter vector $\bm w^{[l]}$ to the $K$ edge nodes, as depicted in Fig. \ref{MB}.

\textbf{Local gradient computation}: Based on the received global model $\bm w^{[l]}$, edge node $k \in {\cal K}$ leverages its own dataset to compute a local gradient vector, given by
\begin{equation*}
    {\bm g}_k^{[l]} = \nabla F_k\left(\bm w^{[l]}\right).
\end{equation*}

\textbf{Edge node selection}: The cloud server selects a subset of the $K$ edge nodes, denoted by ${\cal S}^{[l]} \subseteq {\cal K}$, to participate in the subsequent model aggregation procedure. In the sequel, we term the edge nodes in ${\cal S}^{[l]}$ active edge nodes at the $l$-th training round.

\textbf{Model aggregation}: As shown in Fig. \ref{CMA}, the selected edge nodes upload their respectively computed local gradient vectors to the cloud server, which takes an average of these local gradient vectors to update the global model parameter vector, given by
\begin{eqnarray}
    \bm g^{[l]} = \frac{1}{\left|{\cal S}^{[l]}\right|} \sum\limits_{k \in {\cal S}^{[l]}} {\bm g}_k^{[l]}, \label{Eq-DAG} \\[2ex]
    \bm w^{[l+1]} = \bm w^{[l]} - \eta^{[l]} {\bm g}^{[l]}, \label{Eq-CMU}
\end{eqnarray}
where $\eta^{[l]}$ is the learning rate.

\subsection{Broadband Transmission}
To cope with frequency selective fading and the resultant inter-symbol interference, orthogonal frequency division multiplexing (OFDM) is adopted for uplink transmission from edge nodes to the cloud server. Without loss of generality, we assume that the whole bandwidth is divided into $N$ sub-channels, which are also referred to as sub-carriers.

\subsection{IRS-Assisted Communication}
\begin{figure}
\centering
\includegraphics[width = 5.8cm]{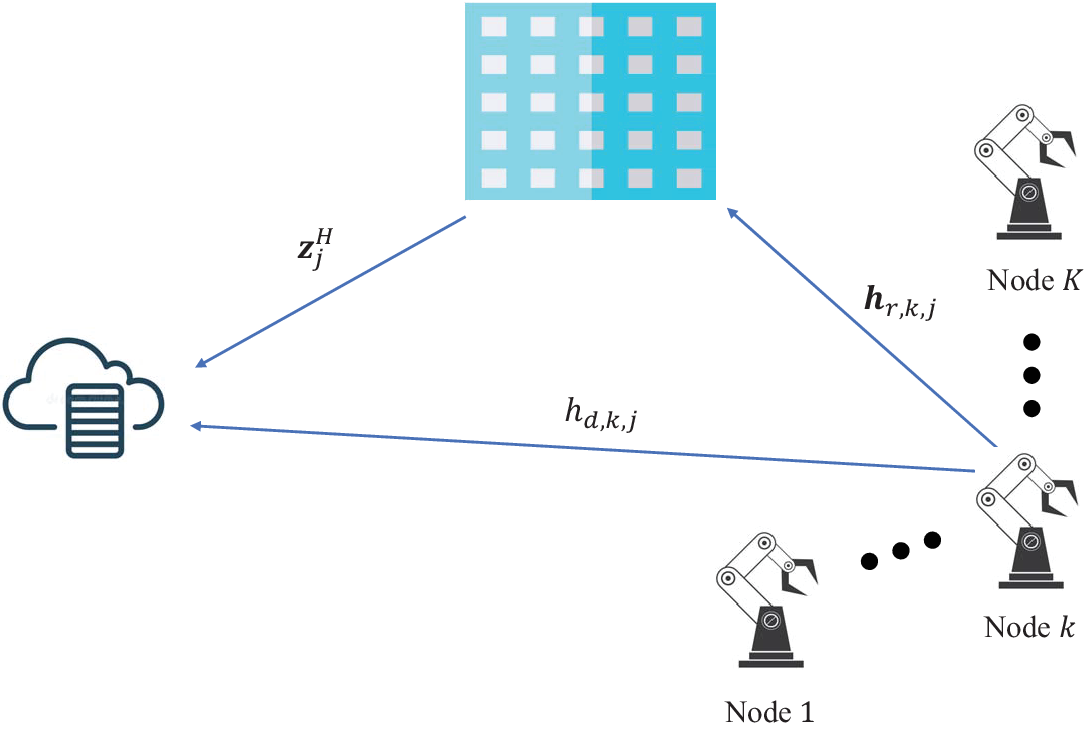}
\caption{An IRS-assisted communication system.} \label{IRS}
\end{figure}

The underlying wireless communication network for the aforementioned FL system is illustrated in Fig. \ref{IRS}, where an IRS is deployed to assist in the communications between the cloud server and the $K$ edge nodes.

Let $h_{d, k, j}$, $\bm h_{r,k,j}$, and $\bm z_j$ denote the $j$-th sub-channel from edge node $k$ to the cloud server, the $j$-th sub-channel from edge node $k$ to the IRS, and the $j$-th sub-channel from the IRS to the cloud server, respectively, $\forall j \in {\cal N} \triangleq \{1, \cdots, N\}$. Moreover, as in \cite{ZhibinWang-TWC}, we denote the diagonal phase-shift matrix of the IRS by $\bm \Theta = \text{diag}\{e^{j \vartheta_1}, \cdots, e^{j \vartheta_P}\}$, where $P$ is the total number of IRS phase shift elements and $\vartheta_p \in [0, 2\pi]$ is the phase shift of element $p$, $\forall p \in {\cal P} \triangleq \{1, \cdots, P\}$. Last, for ease of notation, we define the effective channel from edge node $k$ to the cloud server as
\begin{equation*}
    h_{e,k,j} = h_{d, k, j} + \bm z_j^{H} \bm \Theta \bm h_{r, k, j} = h_{d, k, j} + \bm z_j^{H} \text{diag}\{\bm h_{r, k, j}\} {\bm \theta},
\end{equation*}
where $\bm \theta = [e^{j \vartheta_1}, \cdots, e^{j \vartheta_P}]^{T}$.

\subsection{AirComp-Empowered Model Aggregation}
To reduce the total access delay, we adopt AirComp for model aggregation. That is, every selected edge node occupies all the $N$ sub-channels to upload its computed local gradient parameters to the cloud server in model aggregation. By properly controlling the transmit equalization coefficients of the active edge nodes and the receive factors of the cloud server, a noisy version of the desired global gradient vector \eqref{Eq-DAG} can be constructed, as detailed below.

First of all, we compute the first-order and second-order statistics of each local gradient vector by
\begin{subequations}
    \begin{eqnarray}
    \bar g_k^{[l]} & = & \frac{1}{d} \sum\limits_{i=1}^d \big[\bm g_k^{[l]}\big]_i, \label{First-Order} \\[2ex]
    \delta_k^{[l]} & = & \frac{1}{d} \sum\limits_{i=1}^d \left(\big[\bm g_k^{[l]}\big]_i - \bar g_k^{[l]}\right)^2, \label{Second-Order}
\end{eqnarray}
\end{subequations}
where $\forall k \in {\cal S}^{[l]}$. Then, the two parameters $\bar g_k^{[l]}$ and $\delta_k^{[l]}$ are uploaded to the cloud server waiting for further progressing\footnote{For the sake of simplicity, we follow \cite{HangLiu-TWC} and assume edge node $k$, $\forall k \in {\cal S}^{[l]}$, sends $\bar g_k^{[l]}$ and $\delta_k^{[l]}$ to the cloud server in an error-free fashion.}.

According to \eqref{First-Order} and \eqref{Second-Order}, we can easily derive that
\begin{subequations}
    \begin{eqnarray*}
    && \frac{1}{d} \sum\limits_{i=1}^d \frac{\big[\bm g_k^{[l]}\big]_{i} - \bar g^{[l]}_k}{\sqrt{\delta^{[l]}_k}} = 0, \\[2ex]
    && \frac{1}{d} \sum\limits_{i=1}^d \left(\frac{\big[\bm g_k^{[l]}\big]_{i} - \bar g^{[l]}_k}{\sqrt{\delta^{[l]}_k}}\right)^2 = 1.
\end{eqnarray*}
\end{subequations}
In other words, by using $\bar g_k^{[l]}$ and $\delta_k^{[l]}$, we map $\big[\bm g_k^{[l]}\big]_{i}$ to a zero-mean unit-variance symbol $\frac{\big[\bm g_k^{[l]}\big]_{i} - \bar g^{[l]}_k}{\sqrt{\delta^{[l]}_k}}$, which is the final transmit symbol on the uplink channel.

Recall that there are a total number of $N$ sub-channels that can be leveraged by the active edge nodes to upload their respective local gradient vectors, each of which consists of $d$ parameters. Suppose that each sub-channel conveys only one parameter at one time slot. Then, uploading all $d$ parameters to the cloud server will occupy $T = d/N$ time slots. Particularly, the $\left(j, t\right)$-th received symbol, denoted by $y^{[l]}_{j,t}$, received at the $j$-th sub-channel and $t$-th time slot, is given by
\begin{equation}\label{Eq-RMy}
    y^{[l]}_{j,t} = \sum\limits_{k \in {\cal S}^{[l]}} h^{[l]}_{e,k,j} b^{[l]}_{k,j} \left(\frac{\big[\bm G_k^{[l]}\big]_{j,t} - \bar g^{[l]}_k}{\sqrt{\delta^{[l]}_k}}\right) + n^{[l]}_{j,t},
\end{equation}
where $\bm G^{[l]}_k = \text{mat}(\bm g^{[l]}_k) \in {\cal R}^{N \times T}$, $b^{[l]}_{k,j} \in {\cal C}$ is the transmit equalization coefficient of the $k$-th active edge node at the $j$-th sub-channel, and $n^{[l]}_{j,t} \in {\cal C}$ is the additive white Gaussian noise following ${\cal CN} \left(0, \frac{1}{\textsf{SNR}}\right)$. Note that we consider a block fading channel model in \eqref{Eq-RMy}, where the channel gain coefficient of each link (i.e., $h_{d, k, j}$, $\bm h_{r, k, j}$, and $\bm z_j$) is assumed to be invariant within $O \ge 1$ training rounds\footnote{As reported in \cite{Emil-Book} and \cite{Torres-Electronics}, one coherence block can contain 50000 complex-valued samples in scenarios with low mobility and low channel dispersion, while the dimension of on-device machine learning models is often limited to a few tens of thousands of entries. Therefore, it is considered to be feasible to define the coherence time in terms of the number of FL training rounds.}, such that $h^{[l]}_{e,k,j}$ and $b^{[l]}_{k,j}$ are not related to $t$. Moreover, we follow \cite{ZhibinWang-TWC} and assume perfect channel state information is available at the cloud server, which is in charge of the overall system optimization. Last, the average power constraint for each active node is considered in this paper, such that
\begin{align}\label{Eq-PowerConstraint}
    \frac{1}{d} \sum\limits_{j=1}^N \sum\limits_{t=1}^T \big|b^{[l]}_{k,j}\big|^2 \left(\frac{\big[\bm G_k^{[l]}\big]_{j,t} - \bar g^{[l]}_k}{\sqrt{\delta^{[l]}_k}}\right)^2 ~~~~~~~~~ \nonumber \\[2ex]
    = \frac{1}{N} \sum\limits_{j=1}^N \big|b^{[l]}_{k,j}\big|^2 \tilde \beta^{[l]}_{k,j} ~\le~ 1,
\end{align}
where
\begin{equation*}
    \tilde \beta^{[l]}_{k,j} ~=~ \frac{1}{T} \sum\limits_{t=1}^T \left(\frac{\big[\bm G_k^{[l]}\big]_{j,t} - \bar g^{[l]}_k}{\sqrt{\delta^{[l]}_k}}\right)^2.
\end{equation*}

To successfully update the global model parameter vector, as described in \eqref{Eq-CMU}, the cloud server should be able to recover the global gradient vector from $y^{[l]}_{j,t}$, $\forall j,t$. To achieve this purpose, in addition to controlling the transmit equalization coefficient of each active edge node, the cloud server's receive factors at the $N$ sub-channels, denoted by $m^{[l]}_1, \cdots, m^{[l]}_N \in {\cal C}$ respectively, also remain to be designed. The corresponding signal post-processed by $m_j^{[l]}$ is then given by
\begin{align}\label{Eq-RMr}
    & r^{[l]}_{j,t} = m^{[l]}_j y^{[l]}_{j,t} \nonumber \\[2ex]
    & ~= m^{[l]}_j \sum\limits_{k \in {\cal S}^{[l]}} h^{[l]}_{e,k,j} b^{[l]}_{k,j} \left(\frac{\big[\bm G_k^{[l]}\big]_{j,t} - \bar g^{[l]}_k}{\sqrt{\delta_k^{[l]}}}\right) + m^{[l]}_j n^{[l]}_{j,t}.
\end{align}
Based on \eqref{Eq-RMr}, we can recover a noisy version of the global gradient vector, as detailed below.

Firstly, we set the transmit equalization coefficient $b^{[l]}_{k,j}$ to
\begin{equation}\label{Eq-ZF}
    b^{[l]}_{k,j} = \frac{\sqrt{\delta_k^{[l]}} \left(m_j^{[l]} h^{[l]}_{e,k,j}\right)^{H}}{\left|m_j^{[l]} h^{[l]}_{e,k,j}\right|^2}.
\end{equation}
Given $b^{[l]}_{k,j}$ in \eqref{Eq-ZF}, the power constraint in \eqref{Eq-PowerConstraint} becomes
\begin{equation}\label{Eq-PowerConstraint2}
    \frac{1}{N} \sum\limits_{j=1}^N \big|b^{[l]}_{k,j}\big|^2 \tilde \beta^{[l]}_{k,j} = \frac{1}{N} \sum\limits_{j=1}^N \frac{\beta^{[l]}_{k,j}}{\big|m^{[l]}_j h^{[l]}_{e,k,j} \big|^2} \le 1,
\end{equation}
where $\beta^{[l]}_{k,j} = \tilde \beta_{k, j}^{[l]} \delta_k^{[l]}$. Moreover, the received signal $r^{[l]}_{j,t}$ in \eqref{Eq-RMr} can be rewritten as
\begin{equation}\label{Eq-RMr2}
    r^{[l]}_{j,t} = \sum\limits_{k \in {\cal S}^{[l]}} \left\{\big[\bm G^{[l]}_k\big]_{j,t} - \bar g^{[l]}_k\right\} + m_j^{[l]} n^{[l]}_{j,t}.
\end{equation}
By sequentially executing the following two manipulations: 1) adding $\sum\nolimits_{k \in {\cal S}^{[l]}} \bar g^{[l]}_k$, and 2) multiplying $\frac{1}{|{\cal S}^{[l]}|}$ on both sides of \eqref{Eq-RMr2}, we obtain
\begin{align}\label{Eq-RMr3}
    \big[\hat {\bm G}^{[l]}\big]_{j,t} & = \frac{1}{|{\cal S}^{[l]}|} \left( r^{[l]}_{j,t} + \sum\limits_{k \in {\cal S}^{[l]}} \bar g^{[l]}_k \right) \nonumber \\[2ex]
    & = \frac{1}{|{\cal S}^{[l]}|} \sum\limits_{k \in {\cal S}^{[l]}} \big[\bm G^{[l]}_k\big]_{j,t} + \frac{1}{|{\cal S}^{[l]}|} m_j^{[l]} n^{[l]}_{j,t}.
\end{align}
It can be easily observed that $\big[\hat {\bm G}^{[l]}\big]_{j,t}$ is a noisy version of $\big[\bm G^{[l]}\big]_{j,t} = \frac{1}{|{\cal S}^{[l]}|} \sum\nolimits_{k \in {\cal S}^{[l]}} \big[\bm G^{[l]}_k\big]_{j,t}$. We use MSE to evaluate their difference, given by
\begin{align}\label{Eq-MSE}
    \text{MSE}\left\{\big[\hat {\bm G}^{[l]}\big]_{j,t}, \big[\bm G^{[l]}\big]_{j,t}\right\} & = {\mathbb E}\left[\big|\big[\hat {\bm G}^{[l]}\big]_{j,t} - \big[\bm G^{[l]}\big]_{j,t}\big|^2\right] \nonumber \\[2ex]
    & = \frac{\big|m^{[l]}_j\big|^2}{\textsf{SNR} \left|{\cal S}^{[l]}\right|^2}.
\end{align}
Note that the received global gradient vector through wireless channels inevitably becomes inaccurate due to fading and communication noise\footnote{As in \cite{ZhibinWang-TWC, HangLiu-TWC}, we ignore the errors in the model broadcast procedures and assume that the global model parameter vector is always perfectly received by the edge nodes.}.

\section{Convergence Analysis of Node-Selection Based FL Framework} \label{Section-NSFL-Convergence}
In this section, we analyze the convergence of the node-selection based FL framework that employs IRS, OFDM, and AirComp techniques for model aggregation in each training round.

\subsection{Assumptions}
To proceed, we first make the following assumptions as in \cite{ZhibinWang-TWC, HangLiu-TWC, HaoChen-IoTJ, MPFriedlander}.
\begin{assumption}
\emph{The global loss function $F(\cdot)$ is strongly convex w.r.t. parameter $\mu > 0$, such that for any $\bm x, \bm y \in {\cal R}^d$, we have}
\begin{equation}\label{Eq-ConvexAssump}
    F(\bm y) \ge F(\bm x) + \left(\bm y - \bm x\right)^{T} \nabla F(\bm x) + \frac{\mu}{2} \left\|\bm y - \bm x \right\|^2.
\end{equation}
\end{assumption}
\begin{assumption}
\emph{The global loss function $F(\cdot)$ has Lipschitz continuous gradient with parameter $\rho > 0$, such that for any $\bm x, \bm y \in {\cal R}^d$, we have}
\begin{equation}
    \left\|\nabla F\left(\bm x\right) - \nabla F\left(\bm y\right) \right\| \le \rho \left\|\bm x - \bm y\right\|,
\end{equation}
\emph{which is equivalent to}
\begin{equation}\label{Eq-SmoothAssump}
    F(\bm y) \le F(\bm x) + \left(\bm y - \bm x\right)^{T} \nabla F(\bm x) + \frac{\rho}{2} \left\|\bm y - \bm x \right\|^2.
\end{equation}
\end{assumption}
\begin{assumption}
\emph{The gradient w.r.t. any labeled data sample $\{\bm u_{k,i}, v_{k,i}\}$, $\forall k, i$, is upper bounded. In other words, for some constants $\gamma_1 \ge 0$ and $\gamma_2 \ge 1$, we have}
\begin{equation}\label{Eq-GBAssump}
    \big\|\nabla f\left(\bm w; \bm u_{k,i}, v_{k,i}\right)\big\|^2 \le \gamma_1 + \gamma_2 \big\|\nabla F(\bm w)\big\|^2.
\end{equation}
\end{assumption}

\subsection{Convergence Analysis}
Denote the received noisy global gradient vector by $\hat {\bm g}^{[l]} = \text{vec}\big(\hat {\bm G}^{[l]}\big)$. The global model update recursion in \eqref{Eq-CMU} then becomes
\begin{align}\label{Eq-MDE}
    \bm w^{[l+1]} & = \bm w^{[l]} - \eta^{[l]} \hat {\bm g}^{[l]} \nonumber \\[2ex]
    & = \bm w^{[l]} - \eta^{[l]}\left[\nabla F(\bm w^{[l]}) + \bm e^{[l]}\right].
\end{align}
In \eqref{Eq-MDE}, $\bm e^{[l]}$ is defined as
\begin{align}\label{Eq-ERROR}
    \bm e^{[l]} & = \hat {\bm g}^{[l]} ~-~ \nabla F(\bm w^{[l]}) \nonumber \\[2ex]
    & = \underbrace{\bm g^{[l]} ~-~ \nabla F(\bm w^{[l]})}_{\bm e_1^{[l]}} ~+~ \underbrace{\hat {\bm g}^{[l]} ~-~ \bm g^{[l]}}_{\bm e_2^{[l]}},
\end{align}
where $\bm e_1^{[l]}$ and $\bm e_2^{[l]}$ are respectively attributed to node selection and communication errors. By comparing $\bm g^{[l]}$ with $\nabla F(\bm w^{[l]})$, we can easily observe that $\bm e_1^{[l]} = \bm 0$ when ${\cal S}^{[l]} = {\cal K}$, i.e., all the $K$ edge nodes are selected for model aggregation at the $l$-th training round.

Suppose that the global loss function $F(\bm w)$ indeed satisfies Assumptions 1-3 and the learning rate $\eta^{[l]}$ is set to $\frac{1}{\rho}$. According to \cite{MPFriedlander}, we can obtain that
\begin{equation}\label{Eq-OneIter}
    F(\bm w^{[l+1]}) \le F(\bm w^{[l]}) - \frac{1}{2\rho}\left\|\nabla F(\bm w^{[l]})\right\|^2 + \frac{1}{2\rho} \left\|\bm e^{[l]}\right\|^2.
\end{equation}

In the sequel, by first employing the triangle inequality and then the inequality of arithmetic and geometric means, we upper bound $\big\|\bm e^{[l]}\big\|^2$ as
\begin{align}\label{Eq-ErrorInEq}
    \big\|\bm e^{[l]}\big\|^2 & = \big\|\bm e_1^{[l]} + \bm e_2^{[l]}\big\|^2 \nonumber \\[2ex]
    & \le \left(\big\|\bm e_1^{[l]}\big\| + \big\|\bm e_2^{[l]}\big\|\right)^2 \nonumber \\[2ex]
    & \le 2 \left(\big\|\bm e_1^{[l]}\big\|^2 + \big\|\bm e_2^{[l]}\big\|^2\right).
\end{align}
Following the derivations in Section 3.1 of \cite{MPFriedlander}, we further upper bound $\big\|\bm e_1^{[l]}\big\|^2$ as
\begin{equation}\label{Eq-NSEB}
    \big\|\bm e_1^{[l]}\big\|^2 \le 4 \left(1 - \frac{|{\cal S}^{[l]}|}{K}\right)^2 \left(\gamma_1 + \gamma_2 \left\|\nabla F(\bm w^{[l]})\right\|^2\right).
\end{equation}
Moreover, though $\big\|\bm e_2^{[l]}\big\|^2$ is unbounded, we compute its expectation according to \eqref{Eq-MSE}, given by
\begin{align}\label{Eq-CNEB}
    {\mathbb E}\left[\big\|\bm e_2^{[l]}\big\|^2\right] & = \sum\limits_{j=1}^N \sum\limits_{t=1}^T {\mathbb E}\left[\big|\big[\hat {\bm G}^{[l]}\big]_{j,t} - \big[\bm G^{[l]}\big]_{j,t}\big|^2\right] \nonumber \\[2ex]
    & = \frac{T}{\textsf{SNR} \left|{\cal S}^{[l]}\right|^2} \sum\limits_{j=1}^N \left|m^{[l]}_j\right|^2 \nonumber \\[2ex]
    & \le \frac{d}{\textsf{SNR} \left|{\cal S}^{[l]}\right|^2} \max\limits_{j \in {\cal N}} \left|m^{[l]}_j\right|^2.
\end{align}
Based on \eqref{Eq-OneIter}, \eqref{Eq-ErrorInEq}, \eqref{Eq-NSEB}, and \eqref{Eq-CNEB}, the following theorem can then be derived.
\begin{theorem}\label{Theorem-NSFL-Convergence}
    \emph{Suppose that Assumptions 1-3 are valid and the learning rate is fixed to $\frac{1}{\rho}$. After $L \ge 1$ training rounds, the expected difference between the training loss and the optimal loss can be upper bounded by}
    \begin{align}\label{Eq-NSFL-Convergence}
        {\mathbb E}\left[F(\bm w^{[L+1]}) - F(\bm w^{\star})\right] & \le {\mathbb E}\left[F(\bm w^{[1]}) - F(\bm w^{\star})\right] \prod\limits_{l=1}^L \lambda^{[l]} \nonumber \\[2ex]
        & + \sum\limits_{l=1}^{L} \Psi^{[l]} \prod\limits_{l' = l+1}^L \lambda^{[l']},
    \end{align}
    \emph{where $\bm w^{\star}$ denotes the optimal model parameter vector, and $\lambda^{[l]}$, $\Psi^{[l]}$ are respectively given by}
    \begin{align*}
    \lambda^{[l]} & = 8 \gamma_2 \left(1 - \frac{\left|{\cal S}^{[l]}\right|}{K}\right)^2 + 1 - \frac{\mu}{\rho}, \\[2ex]
    \Psi^{[l]} & = \frac{4 \gamma_1}{\rho} \left(1 - \frac{\left|{\cal S}^{[l]}\right|}{K}\right)^2 + \frac{1}{\rho} \frac{d}{\textsf{SNR} \left|{\cal S}^{[l]}\right|^2} \max\limits_{j \in {\cal N}} \left|m^{[l]}_j\right|^2.
    \end{align*}
\end{theorem}
\begin{IEEEproof}
    Refer to Appendix \ref{Theorem1-Proof}.
\end{IEEEproof}

Moreover, by setting $\left|{\cal S}^{[l]}\right| = K_0$, $\forall l = 1, \cdots, L$, we can simplify \eqref{Eq-NSFL-Convergence} as follows
\begin{align}\label{Eq-NSFL-Convergence2}
    {\mathbb E}\left[F(\bm w^{[L+1]}) - F(\bm w^{\star})\right] ~~~~~~~~~~~~~~~~~~~~~~~~~~~~~ \nonumber \\[2ex]
    \le {\mathbb E}\left[F(\bm w^{[1]}) - F(\bm w^{\star})\right] \lambda_0^L + \sum\limits_{l=1}^L \Psi^{[l]} \lambda_0^{L-l},
\end{align}
where $\lambda_0 = 8 \gamma_2 \left(1 - K_0/ K\right)^2 + 1 - {\mu}/{\rho}$. Suppose $K_0$ is large enough, such that $\lambda_0 < 1$. When $L \rightarrow \infty$, $\lambda_0^L \rightarrow 0$, and therefore, we can further simplify \eqref{Eq-NSFL-Convergence2} as follows
\begin{align}\label{Eq-NSFL-Convergence3}
    & {\mathbb E}\left[F(\bm w^{[L+1]}) - F(\bm w^{\star})\right] \le \sum\limits_{l=1}^L \Psi^{[l]} \lambda_0^{L-l} \\[1ex]
    & = \frac{4 \gamma_1}{\mu \left(1 - K_0 / K\right)^{-2} - 8 \rho \gamma_2} + \sum\limits_{l = 1}^{L} \frac{1}{\textsf{SNR}}\frac{\lambda_0^{L-l} d}{\rho K_0^2} \max\limits_{j \in {\cal N}} \left|m^{[l]}_j\right|^2. \nonumber
\end{align}
It can be seen from \eqref{Eq-NSFL-Convergence3} that the FL recursion is guaranteed to converge with a sufficiently large $K_0$, although there exists a gap between $\lim\nolimits_{L \to \infty}{\mathbb E}\left[F(\bm w^{[L+1]})\right]$ and $F(\bm w^{\star})$ due to node selection, channel fading, and noise.

\subsection{Extension to Non-Convex Loss Function}
Note that the loss function can also be non-convex \cite{ZhaohuiYang-TWC}. In this case, we use the average norm of the global gradient vector to characterize the convergence property of FL recursions, as detailed in the following theorem.
\begin{theorem}\label{Theorem-NonConvex-Convergence}
    \emph{Suppose only Assumptions 2 and 3 are valid and $\eta^{[l]} = \frac{1}{\rho}$, $|{\cal S}^{[l]}| = K_0$, $\forall l = 1, \cdots, L$. After $L$ training rounds, the average norm of the global gradient vector is upper bounded by}
    \begin{align}\label{Eq-NonConvex-Convergence}
        & \frac{1}{L} \sum\limits_{l = 1}^L {\mathbb E}\left[\|\nabla F(\bm w^{[l]})\|^2\right] ~\le~ \frac{2 \rho}{a_0 L}{\mathbb E}\left[F(\bm w^{[1]}) - F(\bm w^{\star})\right] \nonumber \\[2ex]
        & + \left\{\frac{8 \gamma_1}{a_0}\left(1 - \frac{K_0}{K}\right)^2 + \frac{1}{\textsf{SNR}} \frac{2 d}{a_0 L K_0^2}\sum\limits_{l=1}^{L} \max\limits_{j \in {\cal N}} \left|m^{[l]}_j\right|^2\right\},
    \end{align}
    \emph{where $a_0 = 1 - 8 \gamma_2 (1 - K_0/K)^2$.}
\end{theorem}
\begin{IEEEproof}
    Refer to Appendix \ref{Theorem-NonConvex-Proof}.
\end{IEEEproof}
As $L \to \infty$, it is observed that the average norm of the global gradient vector is only determined by the second term on the right side of \eqref{Eq-NonConvex-Convergence}.

\section{Communication and Node Selection Strategy Co-Design }\label{Section-NSFL-PF}
Upon examining \eqref{Eq-NSFL-Convergence3}, we can observe a tradeoff between the node selection loss and the communication error loss, described by the first and second terms of \eqref{Eq-NSFL-Convergence3}, respectively. On the one hand, selecting more edge nodes results in a larger $K_0$, thereby reducing the node selection loss. On the other hand, selecting more edge nodes increases the communication error $\max\nolimits_{j \in {\cal N}} \big|m^{[l]}_j\big|^2$, which will be further elaborated in Proposition \ref{Proposition1}.

In this section, we fix $K_0$ and only seek to minimize the communication errors by jointly optimizing edge node selection, transceiver design, and IRS configuration\footnote{Due to the unknown hyper-parameters, $\mu$, $\rho$, $\gamma_1$, and $\gamma_2$, and the complicated structure of \eqref{Eq-NSFL-Convergence3}, we cannot employ \eqref{Eq-NSFL-Convergence3} as the objective function to construct an optimization problem to optimize $K_0$.}. Below we focus on the $l$-th training round and take the maximum MSE across the $N$ sub-channels of this round as the objective function to construct an optimization problem, given by
\begin{subequations}
    \begin{eqnarray}
        \hspace{-0.7cm} {\textsf P}_1: & \min\limits_{\bm m, \bm \theta, {\cal S}} & \max\limits_{\forall j \in {\cal N}} ~\{|m_j|^2\} \\[1ex]
        \hspace{-0.7cm} & \text{s.t.} & \frac{1}{N} \sum\limits_{j=1}^N \frac{\beta_{k,j}}{\left|m_j h_{e,k,j}(\bm \theta) \right|^2} \le 1, ~\forall k \in {\cal S}, \label{P1-PowerCons} \\[2ex]
        \hspace{-0.7cm} & & \big|[\bm \theta]_p\big| = 1, ~\forall p \in {\cal P}, \label{P1-ThetaCons} \\[2ex]
        \hspace{-0.7cm} & & |{\cal S}| = K_0, \label{P1-NodeCons}
\end{eqnarray}
\end{subequations}
where we have dropped the training round index $l$ for brevity and $\bm m = [m_1, \cdots, m_N]^{T}$.

\begin{proposition}\label{Proposition1}
    \emph{Given $\bm \theta$ and $\cal S$, the optimal receive factors $\bm m^{\star}$ to ${\textsf P}_1$ satisfy the following conditions:}
    \begin{eqnarray}\label{Eq-OptSol}
        |m^\star_1|^2 = \cdots = |m^\star_N|^2 = \max\limits_{\forall k \in {\cal S}} \left\{\frac{1}{N} \sum\limits_{j=1}^N \frac{\beta_{k,j}}{\left|h_{e,k,j}(\bm \theta) \right|^2}\right\}.
    \end{eqnarray}
\end{proposition}
\begin{IEEEproof}
    Refer to Appendix \ref{Proposition1-Proof}.
\end{IEEEproof}

According to this proposition, the optimal value of ${\textsf P}_1$ only depends on the amplitudes of $m_1, \cdots, m_N$, and the phase shifts of these receive factors can be arbitrarily designed. Based on Proposition \ref{Proposition1}, we can reformulate ${\textsf P}_1$ as
\begin{subequations}
    \begin{eqnarray}
        \hspace{-1.1cm} & \min\limits_{\bm \theta, \cal S} & \max\limits_{\forall k \in {\cal S}} \left\{\frac{1}{N} \sum\limits_{j=1}^N \frac{\beta_{k,j}}{\left|h_{e,k,j}(\bm \theta) \right|^2}\right\} \\[2ex]
        \hspace{-1.1cm} & & \big|[\bm \theta]_p\big| = 1, ~\forall p \in {\cal P}, \\[2ex]
        \hspace{-1.1cm} & & |{\cal S}| = K_0,
\end{eqnarray}
\end{subequations}
which is equivalent to
\begin{subequations}
    \begin{eqnarray}
        \hspace{-1.5cm} {\textsf P}_2: & \min\limits_{\bm \theta, \bm \alpha} & \max\limits_{\forall k \in {\cal K}} \left\{\frac{\alpha_k}{N} \sum\limits_{j=1}^N \frac{\beta_{k,j}}{\left|h_{e,k,j}(\bm \theta) \right|^2}\right\} \\[2ex]
        \hspace{-1.5cm} & \text{s.t.} & \big|[\bm \theta]_p\big| = 1, ~~~\forall p \in {\cal P}, \label{P2-ThetaCons} \\[2ex]
        \hspace{-1.5cm} & & \alpha_k \in \{0, 1\}, ~\forall k \in {\cal K}, \label{P2-BinaryCons} \\[1ex]
        \hspace{-1.5cm} & & \sum\limits_{k = 1}^K \alpha_k = K_0, \label{P2-SumCons}
\end{eqnarray}
\end{subequations}
where $\bm \alpha = [\alpha_1, \cdots, \alpha_K]^{T}$ is a binary indicator vector: $\alpha_k = 1$ for $k \in {\cal S}$, and $\alpha_k = 0$ otherwise. Since $\alpha_k \in \{0,1\}$, we can rewrite $\alpha_k$ as $\alpha_k^2$, and transform ${\textsf P}_2$ as follows
\begin{subequations}
    \begin{eqnarray}
        \hspace{-0.7cm} {\textsf P}_3: & \min\limits_{\bm \theta, \bm A, \xi, \bm \alpha} & \xi \\
        \hspace{-0.7cm} & \text{s.t.} & \frac{\alpha_k^2}{N} \sum\limits_{j=1}^N \frac{\beta_{k,j}}{A_{k,j}} \le \xi, ~\forall k \in {\cal K}, \label{P3-GammaCons} \\[1ex]
        \hspace{-0.7cm} & & |h_{e,k,j}(\bm \theta)|^2 = A_{k,j}, \forall k \in {\cal K}, ~\forall j \in {\cal N}, \label{P3-EqualityCons} \\[2ex]
        \hspace{-0.7cm} & & \eqref{P2-ThetaCons}, ~\eqref{P2-BinaryCons}, ~\eqref{P2-SumCons},
\end{eqnarray}
\end{subequations}
where $\xi > 0$ and $\bm A = [A_{1,1}, \cdots, A_{K, N}]$ are introduced auxiliary variables. Due to the non-convexity of \eqref{P2-ThetaCons}, \eqref{P2-BinaryCons} and \eqref{P3-EqualityCons}, it is difficult to solve ${\textsf P}_3$. To address this problem, we leverage the matrix lifting technique and DC representation to transform ${\textsf P}_3$ into a convex problem, as detailed below.

Firstly, we leverage the matrix lifting technique to cope with the non-convexity of \eqref{P2-ThetaCons} and \eqref{P3-EqualityCons}. Note that $|h_{e,k,j}|^2$ can be rewritten as
\begin{eqnarray}\label{Eq-ChannelTransformation}
    |h_{e,k,j}|^2 & = & |h_{d, k, j} + \bm z_j^{H} \text{diag}\{\bm h_{r, k, j}\} {\bm \theta}|^2 \nonumber \\[2ex]
    & = & |h_{d, k, j}|^2 + {\bm \phi}^{H} {\bm H}_{k,j} {\bm \phi} \nonumber \\[2ex]
    & = & |h_{d, k, j}|^2 + \text{Tr}\left(\bm H_{k,j} {\bm \phi} {\bm \phi}^{H}\right).
\end{eqnarray}
In \eqref{Eq-ChannelTransformation}, the two new variables $\bm H_{k,j}$ and ${\bm \phi}$ are respectively defined as
\begin{equation}
    \bm H_{k,j} =
    \begin{bmatrix}
        {\bm h}_{c,k,j} {\bm h}^{H}_{c,k,j} & {\bm h}_{c,k,j} h_{d, k,j} \\[1ex]
        {\bm h}^{H}_{c,k,j} h^{H}_{d,k,j} & 0
    \end{bmatrix}
    ,~
    {\bm \phi} =
    \begin{bmatrix}
        \bm \theta \\[1ex] 1
    \end{bmatrix}
    ,
\end{equation}
where $\bm h^{H}_{c,k,j} = \bm z_j^{H} \text{diag}\{\bm h_{r, k, j}\}$. Next, we introduce $\bm \Phi = \bm \phi \bm \phi^{H}$, and transform $\textsf{P}_3$ as follows
\begin{subequations}
    \begin{align}
        {\textsf P}_4: \min\limits_{\bm \Phi, \bm A, \xi, \bm \alpha}~& \xi \\
        \text{s.t.}~~~~& \text{Tr}(\bm H_{k,j} {\bm \Phi}) + |h_{d, k, j}|^2 = A_{k,j}, \forall k \in {\cal K}, j \in {\cal N}, \label{P4-EqualityCons} \\[1ex]
        & [\bm \Phi]_{n, n} = 1, \forall n \in \{1, \cdots, P + 1\}, \label{P4-PhiCons} \\[1ex]
        & \text{Rank}(\bm \Phi) = 1, \\[1ex]
        & \eqref{P2-BinaryCons}, ~\eqref{P2-SumCons}, ~\eqref{P3-GammaCons}.
\end{align}
\end{subequations}
In addition, $\text{Rank}(\bm \Phi) = 1$ and $\alpha_k \in \{0, 1\}$ can be equivalently rewritten as
\begin{subequations}
    \begin{eqnarray}
        & & \text{Tr}(\bm \Phi) - \|\bm \Phi\| = 0, \\[1ex]
        & & \alpha_k \in [0, 1],~\alpha_k (1 - \alpha_k) \le 0.
\end{eqnarray}
\end{subequations}
Consequently, we can further transform $\textsf{P}_4$ as
\begin{subequations}
    \begin{align}
        {\textsf P}_5: \min\limits_{\bm \Phi, \bm A, \xi, \bm \alpha}~& \xi + \Delta_1 (\text{Tr}(\bm \Phi) - \|\bm \Phi\|) + \Delta_2 (\bm 1^{T} \bm \alpha - \|\bm \alpha\|^2) \label{P5-Obj} \\[1ex]
        \text{s.t.}~~~~& 0 \le \alpha_k \le 1, \forall k \in {\cal K}, \label{P5-AlphaCons} \\[1ex]
        & \eqref{P2-SumCons}, ~\eqref{P3-GammaCons}, ~\eqref{P4-EqualityCons}, ~\eqref{P4-PhiCons},
\end{align}
\end{subequations}
where $\Delta_1 > 0$ and $\Delta_2 > 0$ are two penalty parameters. While the objective function of ${\textsf P}_5$ is still non-convex, its structure of minimizing the difference between two convex functions can be leveraged to develop efficient DC algorithms. In what follows, we use the successive convex approximation (SCA) technique to solve ${\textsf P}_5$. Specifically, at iteration $n + 1$, by linearizing the concave parts in \eqref{P5-Obj}, i.e.,
\begin{align*}
    \|\bm \Phi\|~ & \ge \|\bm \Phi_n\| + \langle \partial_{\bm \Phi_n} \|\bm \Phi\|, \bm \Phi - {\bm \Phi_n} \rangle = \text{Tr}(\bm \omega_n \bm \omega_n^{H} \bm \Phi), \\[2ex]
    \|\bm \alpha\|^2 & \ge \|\bm \alpha_n\|^2 + 2 \bm \alpha_n^{T}(\bm \alpha - \bm \alpha_n),
\end{align*}
we construct a convex optimization problem given by
\begin{subequations}
    \begin{align*}
        {\textsf P}_6: \min\limits_{\bm \Phi, \bm A, \xi, \bm \alpha}~& \xi + \Delta_1 \text{Tr}((\bm I - \bm \omega_n \bm \omega_n^{H}) \bm \Phi) \\[1ex]
        &~~~~+ \Delta_2 [\|\bm \alpha_n\|^2 + (\bm 1 - 2 \bm \alpha_n)^{T} \bm \alpha] \\[2ex]
        \text{s.t.}~~~~& \eqref{P2-SumCons}, ~\eqref{P3-GammaCons}, ~\eqref{P4-EqualityCons}, ~\eqref{P4-PhiCons}, ~\eqref{P5-AlphaCons},
\end{align*}
\end{subequations}
where $\langle \cdot, \cdot \rangle$ defines the inner product of two matrices, and $\partial_{\bm \Phi_n} \|\bm \Phi\| = \bm \omega_n \bm \omega_n^{H}$ denotes the sub-gradient of $\|\bm \Phi\|$ at $\bm \Phi_n$, and $\bm \omega_n$ is the eigenvector corresponding to the largest eigenvalue of $\bm \Phi_n$. By applying the same procedures employed in proving Proposition 5 of \cite{YuanmingShi-TWC}, we can establish that the solution sequence generated by recursively solving ${\textsf P}_6$ always converges to a critical point of ${\textsf P}_5$ given an arbitrary initial point.

Solving ${\textsf P}_6$ successively until convergence, we obtain a rank-one $\bm \Phi$ when $\text{Tr}(\bm \Phi) - \|\bm \Phi\| = 0$, and a $0-1$ vector $\bm \alpha$ when $\bm 1^{T} \bm \alpha - \|\bm \alpha\|^2 = 0$, denoted by $\bm \Phi^{\star}$ and $\bm \alpha^{\star}$, respectively. We then extract ${\bm \phi}^{\star}$ by doing Cholesky decomposition for $\bm \Phi^{\star}$, and the phase shift vector of the IRS is given by $\bm \theta^{\star} = {[{\bm \phi}^{\star}]_{1:P}}/{[{\bm \phi}^{\star}]_{P+1}}$. Finally, since the computational cost of solving ${\textsf P}_6$ via the second-order interior point method \cite{Boyd-Book} is ${\cal O}([(P+1)^2 + K]^3)$ at each iteration, the overall computational cost of solving ${\textsf P}_3$ is ${\cal O}(I_1[(P+1)^2 + K]^3)$, where $I_1$ is the total number of iterations before ${\textsf P}_6$ converges.

Until now, we have introduced how to select edge nodes, design their transmit equalization coefficients, and optimize the receive factors of the cloud server and the IRS phase shifts to control the communication errors in model aggregation. In the next section, we propose a new FL framework to improve learning performance further.

\section{Weight-Selection Based FL Systems}\label{Section-WSFL}
As mentioned before, although node selection helps mitigate aggregation errors, dropping a few edge nodes can lead to a non-negligible decrease in the learning performance, especially when the discarded edge nodes have unique data samples.

To avoid this dilemma, instead of dropping any edge nodes in model aggregation, we assign each edge node a carefully designed weight coefficient to control the aggregation error, as shown in Fig. \ref{PMA}. In other words, instead of utilizing \eqref{Eq-CMU}, we propose to update the global model by
\begin{equation}\label{Eq-PMU}
    \bm w^{[l+1]} = \bm w^{[l]} - \frac{\eta^{[l]}}{\sum\nolimits_{j=1}^K q^{[l]}_j} \sum\limits_{k=1}^K q_k^{[l]} {\bm g}_k^{[l]},
\end{equation}
where $q_k^{[l]} \in \left[0,1\right]$ is the $k$-th edge node's weight coefficient for model aggregation at the $l$-th training round, $\forall k \in {\cal K}$. Notice that \eqref{Eq-PMU} can be viewed as a generalization of \eqref{Eq-CMU}, and if $q_1^{[l]}, \cdots, q_K^{[l]}$ are restricted to be $0-1$ binary variables, \eqref{Eq-PMU} reduces to \eqref{Eq-CMU}. In the following, we first introduce how to recover
\begin{equation}\label{Eq-PDGG}
    \bm{\tilde g}^{[l]} = \frac{1}{\sum\nolimits_{j=1}^K q^{[l]}_j} \sum\limits_{k=1}^K q_k^{[l]} {\bm g}_k^{[l]}
\end{equation}
from the received signals, and then analyze the performance of such a weight-selection based FL system. At the end of this section, we discuss how to design $q_1^{[l]}, \cdots, q_K^{[l]}$ to achieve appealing learning performance.

\subsection{Reconstruction of the Desired Global Gradient Vector}
To recover \eqref{Eq-PDGG} from $r_{j,t}^{[l]}$, we set the transmit equalization coefficient $b^{[l]}_{k,j}$ to
\begin{equation}\label{Eq-PZF}
    b^{[l]}_{k,j} = \frac{\sqrt{\delta_k^{[l]}} q_k^{[l]} \left(m_j^{[l]} h^{[l]}_{e,k,j}\right)^{H}}{\left|m_j^{[l]} h^{[l]}_{e,k,j}\right|^2}.
\end{equation}
Given $b^{[l]}_{k,j}$ in \eqref{Eq-PZF}, the power constraint in \eqref{Eq-PowerConstraint} now becomes
\begin{equation}\label{Eq-PowerConstraint3}
    \frac{1}{N} \sum\limits_{j=1}^N \big|b^{[l]}_{k,j}\big|^2 \tilde \beta^{[l]}_{k,j} = \frac{1}{N} \sum\limits_{j=1}^N \frac{\beta^{[l]}_{k,j} \big(q^{[l]}_k\big)^2} {\big|m^{[l]}_j h^{[l]}_{e,k,j} \big|^2} \le 1.
\end{equation}
Then, the received signal $r^{[l]}_{j,t}$ in \eqref{Eq-RMr} can be rewritten as
\begin{equation}\label{Eq-PRMr2}
    r^{[l]}_{j,t} = \sum\limits_{k = 1}^K q^{[l]}_k \left\{\big[\bm G^{[l]}_k\big]_{j,t} - \bar g^{[l]}_k\right\} + m_j^{[l]} n^{[l]}_{j,t}.
\end{equation}
By sequentially executing the following two manipulations: 1) adding $\sum\nolimits_{k = 1}^K q_k^{[l]} \bar g_k^{[l]}$, and 2) multiplying $\frac{1}{Q^{[l]}}$ on both sides of \eqref{Eq-PRMr2}, we obtain
\begin{align}\label{Eq-PRMr3}
    \big[\bm{\breve G}^{[l]}\big]_{j,t} & = \frac{1}{Q^{[l]}} \left(r^{[l]}_{j,t} + \sum\limits_{k = 1}^K q_k^{[l]} \bar g^{[l]}_k \right) \nonumber \\[2ex]
    & = \frac{1}{Q^{[l]}} \sum\limits_{k = 1}^K q_k^{[l]} \big[\bm G^{[l]}_k\big]_{j,t} + \frac{1}{Q^{[l]}} m_j^{[l]} n^{[l]}_{j,t},
\end{align}
where $Q^{[l]} = \sum\nolimits_{k = 1}^K q_k^{[l]}$. It can be observed that $\big[\bm{\breve G}^{[l]}\big]_{j,t}$ is a noisy version of $\big[\bm{\tilde G}^{[l]}\big]_{j,t} = \frac{1}{Q^{[l]}} \sum\nolimits_{k = 1}^K q_k^{[l]} \big[\bm G^{[l]}_k\big]_{j,t}$, where $\bm{\tilde G} = \text{mat}(\bm{\tilde g})$. We continue using MSE to evaluate the difference between $\big[\bm{\breve G}^{[l]}\big]_{j,t}$ and $\big[\bm{\tilde G}^{[l]}\big]_{j,t}$, given by
\begin{align}\label{Eq-PMSE}
    \text{MSE}\left\{\big[\bm{\breve G}^{[l]}\big]_{j,t}, \big[\bm{\tilde G}^{[l]}\big]_{j,t}\right\} & = {\mathbb E}\left[\big|\big[\bm{\breve G}^{[l]}\big]_{j,t} - \big[\bm{\tilde G}^{[l]}\big]_{j,t}\big|^2\right] \nonumber \\[2ex]
    & = \frac{\big|m^{[l]}_j\big|^2}{\textsf{SNR} \left(Q^{[l]}\right)^2}.
\end{align}

\subsection{Convergence Analysis}
Following \eqref{Eq-ERROR}, we also decompose the gradient error into two parts, i.e.,
\begin{equation}\label{Eq-PGEV}
    \bm{\tilde e}^{[l]} ~=~ \underbrace{\bm{\tilde g}^{[l]} ~-~ \nabla F(\bm w^{[l]})}_{\bm{\tilde e}_1^{[l]}} ~+~ \underbrace{\bm{\breve g}^{[l]} ~-~ \bm{\tilde g}^{[l]}}_{\bm{\tilde e}_2^{[l]}},
\end{equation}
where $\bm {\tilde e}_1$ and $\bm {\tilde e}_2$ are respectively attributed to weight selection and communication errors, and $\bm{\breve g} = \text{vec}(\bm{\breve G})$. By comparing $\bm{\tilde g}^{[l]} = \frac{1}{Q^{[l]}} \sum\nolimits_{k=1}^K q_k^{[l]} {\bm g}_k^{[l]}$ with $\nabla F(\bm w^{[l]}) = \frac{1}{K} \sum\nolimits_{k=1}^K {\bm g}_k^{[l]}$, we can easily observe that $\bm{\tilde e}_1^{[l]} = \bm 0$ when $q^{[l]}_1 = \cdots = q^{[l]}_K = 1$.

By resorting to \eqref{Eq-PMSE}, we can derive that
\begin{align}\label{Eq-PNEB}
    {\mathbb E}\left[\big\|\bm{\tilde e}_2^{[l]}\big\|^2\right] & = \sum\limits_{j=1}^N \sum\limits_{t=1}^T {\mathbb E}\left[\big|\big[\bm{\breve G}^{[l]}\big]_{j,t} - \big[\bm{\tilde G}^{[l]}\big]_{j,t}\big|^2\right] \nonumber \\[2ex]
    & = \frac{T}{\textsf{SNR} \left(Q^{[l]}\right)^2} \sum\limits_{j=1}^N \left|m^{[l]}_j\right|^2 \nonumber \\[2ex]
    & \le \frac{d}{\textsf{SNR} \left(Q^{[l]}\right)^2} \max\limits_{j \in {\cal N}} \left|m^{[l]}_j\right|^2.
\end{align}
As for $\bm{\tilde e}_1^{[l]}$, we provide an upper bound for $\big\|\bm{\tilde e}_1^{[l]}\big\|^2$ in the following lemma.
\begin{lemma}\label{Lemma-Bound}
    \emph{The error due to weight selection can be upper bounded by}
    \begin{equation}\label{Eq-SelErrUB}
        \big\|\bm{\tilde e}_1^{[l]}\big\|^2 \le 4 \left(1 - \frac{Q^{[l]}}{K}\right)^2 \left(\gamma_1 + \gamma_2 \left\|\nabla F(\bm w^{[l]})\right\|^2\right).
    \end{equation}
\end{lemma}
\begin{IEEEproof}
Given a random mini-batch $\widetilde{\cal D}^{[l]}_k \subseteq {\cal D}_k$, whose cardinality is denoted by $\widetilde D^{[l]}_k$, $\forall k \in {\cal K}$, we introduce an auxiliary variable $\bm{\breve e}_1^{[l]}$, defined as
\begin{align}\label{Eq-AuxErrorExpression}
    \bm{\breve e}_1^{[l]} & = \frac{1}{\sum\nolimits_{j=1}^K \widetilde D^{[l]}_j} \sum\limits_{k=1}^K \sum\limits_{i = 1}^{\widetilde D^{[l]}_k} \nabla f\left(\bm w^{[l]}; \tilde{\bm u}_{k, i}, \tilde v_{k,i}\right) \nonumber \\[2ex]
    & - \frac{1}{KD} \sum\limits_{k=1}^K \sum\limits_{i=1}^{D} \nabla f\left(\bm w^{[l]}; \bm u_{k,i}, v_{k,i}\right),
\end{align}
where $\{\tilde{\bm u}_{k, i}, \tilde{v}_{k,i}\} \in \widetilde {\cal D}^{[l]}_k$. According to \cite{MiniBatch-SGD}, we have
\begin{align}\label{Eq-UBGrad}
    {\mathbb E}\left[\sum\limits_{i = 1}^{\widetilde D^{[l]}_k} \nabla f\left(\bm w^{[l]}; \tilde {\bm u}_{k,i}, \tilde v_{k,i}\right)\right] ~~~~~~~~~~~~~~~~~~ \nonumber \\[2ex]
    = \frac{\widetilde D^{[l]}_k}{D} \sum\limits_{i = 1}^D \nabla f\left(\bm w^{[l]}; \bm u_{k,i}, v_{k,i}\right).
\end{align}
In \eqref{Eq-UBGrad}, the expectation is taken over $\widetilde {\cal D}^{[l]}_k$, $\forall k \in {\cal K}$. Recall that
\begin{align}\label{Eq-ErrorExpression}
    \bm{\tilde e}_1^{[l]} & = \bm{\tilde g}^{[l]} - \nabla F\left(\bm w^{[l}\right) \nonumber \\[2ex]
    & = \frac{1}{Q^{[l]} D} \sum\limits_{k=1}^K q_k^{[l]} \sum\limits_{i=1}^{D} \nabla f\left(\bm w^{[l]}; \bm u_{k,i}, v_{k,i}\right) \nonumber \\[2ex]
    & - \frac{1}{KD} \sum\limits_{k=1}^K \sum\limits_{i=1}^{D} \nabla f\left(\bm w^{[l]}; \bm u_{k,i}, v_{k,i}\right).
\end{align}
By setting $\widetilde D_k^{[l]} = q_k^{[l]} D$, $\forall k \in {\cal K}$, we can derive that
\begin{equation}\label{Eq-UBError}
    {\mathbb E}\big[\bm{\breve e}_1^{[l]}\big] ~=~ \bm{\tilde e}_1^{[l]},
\end{equation}
where the expectation is taken over $\widetilde{\cal D}_1^{[l]}, \cdots, \widetilde{\cal D}_K^{[l]}$.

According to \cite{MPFriedlander}, since $\big\|\bm{\breve e}_1^{[l]}\big\|^2$ can be upper bounded by
\begin{align*}
    \big\|\bm{\breve e}_1^{[l]}\big\|^2 & \le 4 \left(1 - \frac{\sum\nolimits_{k=1}^K \widetilde D_k^{[l]}}{KD}\right)^2 \left(\gamma_1 + \gamma_2 \left\|\nabla F\left(\bm w^{[l]}\right)\right\|^2\right) \\[2ex]
    & = 4 \left(1 - \frac{Q^{[l]}}{K}\right)^2 \left(\gamma_1 + \gamma_2 \left\|\nabla F\left(\bm w^{[l]}\right)\right\|^2\right),
\end{align*}
we thus derive that
\begin{align}\label{Eq-SelErrUBProof}
    \big\|\bm{\tilde e}_1^{[l]}\big\| & = \big\|{\mathbb E}\big[\bm{\breve e}_1^{[l]}\big]\big\| \overset{(a)}{\le} {\mathbb E}\big[\big\|\bm{\breve e}_1^{[l]}\big\|\big] \nonumber \\[2ex]
    & \le 2 \left(1 - \frac{ Q^{[l]}}{K}\right) \sqrt{\gamma_1 + \gamma_2 \left\|\nabla F\left(\bm w^{[l]}\right)\right\|^2},
\end{align}
where $(a)$ is due to the convexity of $\ell_2$-norm. Until now, we have proven Lemma \ref{Lemma-Bound}.
\end{IEEEproof}

Based on \eqref{Eq-UBError}, we view $\bm{\breve e}_1^{[l]}$ as a surrogate variable of $\bm{\tilde e}_1^{[l]}$. Moreover, since $\bm{\breve e}_1^{[l]}$ can be explained as the gradient error incurred by FL training with $\widetilde{\cal D}_1^{[l]}, \cdots, \widetilde {\cal D}_K^{[l]}$, and $\widetilde D_k^{[l]} = q_k^{[l]} D$, $\forall k \in {\cal K}$, we thus term $q_k^{[l]} \in \left[0, 1\right]$ the ratio of the number of selected samples to the total number of samples in ${\cal D}_k$ for FL training at the $l$-th training round, $\forall k \in {\cal K}$. The convergence performance of such a weight-selection based FL system is provided in the following theorem.
\begin{theorem}\label{Theoren-WSFL-Convergence}
    \emph{Suppose that Assumptions 1-3 are valid and the learning rate is fixed to $\frac{1}{\rho}$. After $L \ge 1$ training rounds, the expected difference between the training loss and the optimal loss can be upper bounded by}
    \begin{align}\label{Eq-WSFL-Convergence}
        {\mathbb E}\left[F(\bm w^{[L+1]}) - F(\bm w^{\star})\right] & \le {\mathbb E}\left[F(\bm w^{[1]}) - F(\bm w^{\star})\right] \prod\limits_{l=1}^L \tilde \lambda^{[l]} \nonumber \\[2ex]
        & + \sum\limits_{l=1}^{L} \tilde \Psi^{[l]} \prod\limits_{l' = l+1}^L \tilde \lambda^{[l']},
    \end{align}
    \emph{where $\tilde \lambda^{[l]}$ and $\tilde \Psi^{[l]}$ are respectively given by}
    \begin{align*}
        \tilde \lambda^{[l]} & = 8 \gamma_2 \left(1 - \frac{Q^{[l]}}{K}\right)^2 + 1 - \frac{\mu}{\rho}, \\[2ex]
        \tilde \Psi^{[l]} & = \frac{4 \gamma_1}{\rho} \left(1 - \frac{Q^{[l]}}{K}\right)^2 + \frac{1}{\rho} \frac{d}{\textsf{SNR} \left(Q^{[l]}\right)^2} \max\limits_{j \in {\cal N}} \left|m^{[l]}_j\right|^2.
    \end{align*}
\end{theorem}
\begin{IEEEproof}
    The proof of this theorem is the same as that of Theorem \ref{Theorem-NSFL-Convergence} except for replacing ${\cal S}^{[l]}$ with $Q^{[l]}$, and hence is omitted for brevity.
\end{IEEEproof}

By setting $Q^{[l]} = Q_0$, $\forall l = 1, \cdots, L$, we can simplify \eqref{Eq-WSFL-Convergence} as follows
\begin{align}\label{Eq-WSFL-Convergence2}
    {\mathbb E}\left[F(\bm w^{[L+1]}) - F(\bm w^{\star})\right] ~~~~~~~~~~~~~~~~~~~~~~~~~~~~~ \nonumber \\[2ex]
    \le {\mathbb E}\left[F(\bm w^{[1]}) - F(\bm w^{\star})\right] \tilde \lambda_0^L + \sum\limits_{l=1}^L \tilde \Psi^{[l]} \tilde \lambda_0^{L-l},
\end{align}
where $\tilde \lambda_0 = 1 - {\mu}/{\rho} + 8 \gamma_2 \left(1 - Q_0 / K\right)^2$. Suppose $Q_0$ is large enough, such that $\tilde \lambda_0 < 1$. When $L \to \infty$, $\tilde \lambda_0 \to 0$, and we can further simplify \eqref{Eq-WSFL-Convergence2} as follows
\begin{align}\label{Eq-WSFL-Convergence3}
    & {\mathbb E}\left[F(\bm w^{[L+1]}) - F(\bm w^{\star})\right] \le \sum\limits_{l=1}^L \tilde \Psi^{[l]} \tilde \lambda_0^{L-l} \\[1ex]
    & = \frac{4 \gamma_1}{\mu \left(1 - Q_0 / K\right)^{-2} - 8 \rho \gamma_2} + \sum\limits_{l = 1}^{L} \frac{1}{\textsf{SNR}} \frac{\tilde \lambda_0^{L-l} d}{\rho Q_0^2} \max\limits_{j \in {\cal N}} \left|m^{[l]}_j\right|^2. \nonumber
\end{align}

\subsection{Weight Coefficient Optimization}
Similar to \eqref{Eq-NSFL-Convergence3}, we observe a tradeoff between the weight selection loss and the communication error loss, described by the first and second terms of \eqref{Eq-WSFL-Convergence3}, respectively. As in Section \ref{Section-NSFL-PF}, we fix the weight selection loss by setting $\sum\nolimits_{k = 1}^K q_k^{[l]}$ to $Q_0$, $\forall l = 1, \cdots, L$, and only seek to minimize the communication errors by optimizing $q_1^{[l]}, \cdots, q_K^{[l]}$. Below we focus on the $l$-th training round and take the maximum MSE across the $N$ sub-channels of this round as the objective function to construct an optimization problem, given by
\begin{subequations}
    \begin{eqnarray}
        \hspace{-0.7cm} {\textsf P}_7: & \min\limits_{\bm q, \bm m, \bm \theta} & \max\limits_{\forall j \in {\cal N}} ~\{|m_j|^2\} \\[1ex]
        \hspace{-0.7cm} & \text{s.t.} & \frac{1}{N} \sum\limits_{j=1}^N \frac{\beta_{k,j} q_k^2}{\left|m_j h_{e,k,j}(\bm \theta) \right|^2} \le 1, ~\forall k \in {\cal K}, \label{P7-PowerCons} \\[2ex]
        \hspace{-0.7cm} & & \big|[\bm \theta]_p\big| = 1, ~\forall p \in {\cal P}, \label{P7-ThetaCons} \\[2ex]
        \hspace{-0.7cm} & & q_k \in [0, 1], ~\forall k \in {\cal K}, \label{P7-qbound} \\[2ex]
        \hspace{-0.7cm} & & \sum\limits_{k = 1}^K q_k^{[l]} = Q_0, \label{P7-qsum}
\end{eqnarray}
\end{subequations}
where we have dropped the training round index $l$ for convenience and $\bm q = [q_1, \cdots, q_K]^{T}$.

\begin{proposition}\label{Proposition2}
    \emph{Given $\bm q$ and $\bm \theta$, the optimal receive factors $\bm m^{\star}$ to ${\textsf P}_7$ satisfy the following conditions:}
    \begin{eqnarray}\label{Eq-WSFL-OptSol}
        |m^\star_1|^2 = \cdots = |m^\star_N|^2 = \max\limits_{\forall k \in {\cal K}} \left\{\frac{q_k^2}{N} \sum\limits_{j=1}^N \frac{\beta_{k,j}}{\left|h_{e,k,j}(\bm \theta) \right|^2}\right\}.
    \end{eqnarray}
\end{proposition}
\begin{IEEEproof}
    The proof of this proposition is the same as that of Proposition \ref{Proposition1} except for replacing (\ref{P1-PowerCons}) with (\ref{P7-PowerCons}), and thus is omitted.
\end{IEEEproof}

Based on Proposition \ref{Proposition2}, we can transform ${\textsf P}_7$ into
\begin{subequations}
    \begin{eqnarray}
        \hspace{-0.7cm} & \min\limits_{\bm q, \bm \theta} & \max\limits_{\forall k \in {\cal K}} \left\{\frac{q_k^2}{N} \sum\limits_{j=1}^N \frac{\beta_{k,j}}{\left|h_{e,k,j}(\bm \theta) \right|^2}\right\} \\[2ex]
        \hspace{-0.7cm} & \text{s.t.} & \eqref{P7-ThetaCons}, ~\eqref{P7-qbound}, ~\eqref{P7-qsum},
\end{eqnarray}
\end{subequations}
which can be reformulated as
\begin{subequations}
    \begin{eqnarray}
        \hspace{-0.7cm} {\textsf P}_8: & \min\limits_{\bm \theta, \bm A, \xi, \bm q} & \xi \\
        \hspace{-0.7cm} & \text{s.t.} & \frac{q_k^2}{N} \sum\limits_{j=1}^N \frac{\beta_{k,j}}{A_{k,j}} \le \xi, ~\forall k \in {\cal K}, \\[2ex]
        \hspace{-0.7cm} & & \left|h_{e,k,j}(\bm \theta) \right|^2 = A_{k,j}, ~\forall k \in {\cal K}, ~\forall j \in {\cal N}, \\[2ex]
        \hspace{-0.7cm} & & \eqref{P7-ThetaCons}, ~\eqref{P7-qbound}, ~\eqref{P7-qsum}.
\end{eqnarray}
\end{subequations}
It is observed that ${\textsf P}_8$ is almost the same as $\textsf{P}_3$ except for replacing $\bm \alpha$ with $\bm q$. Therefore, we can similarly leverage the matrix lifting technique and DC programming to solve ${\textsf P}_8$, but for brevity, we omit the details here. Since $\bm q$ is a real vector while $\bm \alpha$ is only a binary one, ${\textsf P}_8$ will achieve a lower objective value than ${\textsf P}_3$, and we can thus infer that the weight-selection based FL framework will outperform its node-selection based counterpart.

Thus far, we have introduced the weight-selection based FL framework. We have described how to design the edge nodes' transmit equalization coefficients, their weight coefficients, the cloud server's receive factors, and the IRS phase shifts to control the communication errors in model aggregation. In the next section, we use experiments to evaluate the performance of the node-selection and weight-selection based FL frameworks.

\section{Numerical Results} \label{Section-Results}
We consider a three-dimensional coordinate system, where the locations of the cloud server and the IRS are respectively set to $\left(-50, 0, 10\right)$ meters and $\left(0, 0, 10\right)$ meters, and the $K = 20$ edge nodes are uniformly distributed in the region of $\left([0,20], [-10, 10], 0\right)$ meters. The whole bandwidth is divided into $N = 10$ sub-channels, and $\{h_{d,k,j}\}$, $\{\bm h_{r,k,j}\}$, and $\{\bm z_j\}$ suffer from both path loss and small scale fading. The path loss model is expressed as $\textsf{PL}(\zeta) = C_0 \left(\zeta / \zeta_0\right)^{-\kappa}$, where $C_0 = 30$ dB accounts for the path loss at the reference distance of $\zeta_0 = 1$ meter, $\zeta$ denotes the link distance, and $\kappa$ is the path loss component. Following \cite{ZhibinWang-TWC}, the path loss components for $\{h_{d,k,j}\}$, $\{\bm h_{r,k,j}\}$, and $\{\bm z_j\}$, are respectively set to 3.6, 2.8 and 2.2. The small-scale fading coefficients are assumed to follow the Rician distribution, given by
\begin{equation*}
    \sqrt{\frac{\chi}{1 + \chi}} {\bm 1} + \sqrt{\frac{1}{1 + \chi}}{\cal CN}(\bm 0, \bm I),
\end{equation*}
where $\chi$ is termed the Rician factor. As in \cite{ZhibinWang-TWC}, the Rician factors for $\{h_{d,k,j}\}$, $\{\bm h_{r,k,j}\}$, and $\{\bm z_j\}$ are respectively set to 0, 0, and 3 dB. Moreover, we set $\textsf{SNR} = 60$ dB, $P = 20$, and $O = 1$ unless otherwise specified.

\begin{figure}
\centering
\includegraphics[width = 8.6cm]{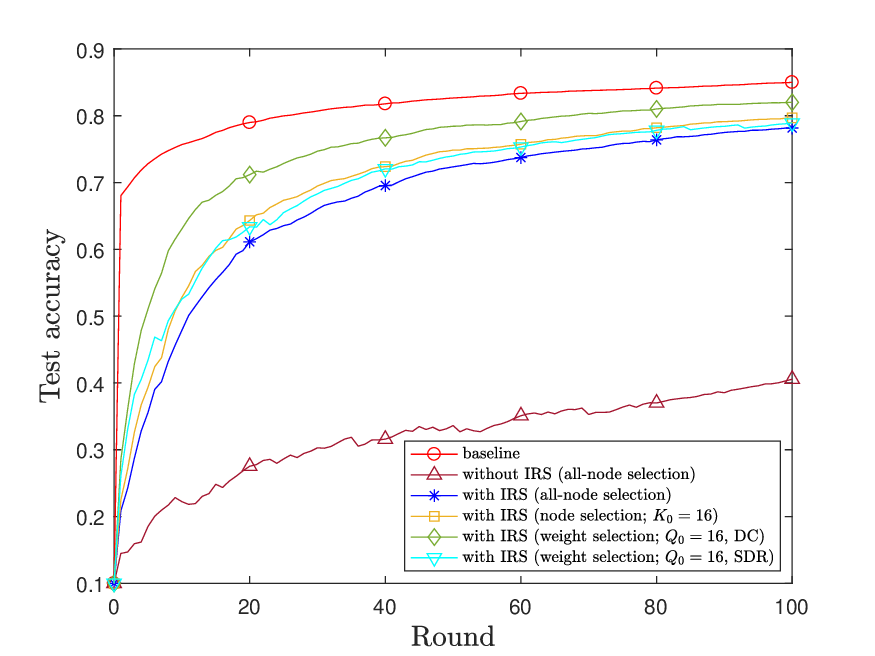}
\caption{Test accuracy w.r.t. training round.} \label{TA-TrainingRound}
\end{figure}

In regards to the learning purpose, we use the MNIST dataset \cite{MNIST} to simulate the handwritten digit recognition task. Specifically, by using cross-entropy as the loss function, we train a fully connected neural network consisting of 784 inputs and 10 outputs, i.e., the number of model parameters $d = 7840$. The training set of 60,000 samples is equally divided into 20 shards of size $D = 3000$ in a non-IID manner, and each shard is assigned to one edge node as its local dataset. The test dataset has 10,000 different samples and we adopt test accuracy, defined as $\frac{\text{\# of correctly recognized handwritten-digits}}{10000} \in [0,1]$, to evaluate the FL learning performance. The total number of training rounds $L$ is set to 100, and the learning rate $\eta^{[l]} = 0.01$, $\forall l \in \{1, \cdots, L\}$. Furthermore, the following baselines are used for comparison in the simulations.

\textbf{1) Ideal baseline}: All the $K$ edge nodes are selected with $q^{[l]}_k = 1$, $\forall k \in {\cal K}$, and the cloud server receives $\frac{1}{K} \sum\nolimits_{k=1}^K {\bm g}^{[l]}_k$, in an error-free manner.

\textbf{2) All-node selection}: All the $K$ edge nodes are selected with $q^{[l]}_k = 1$, $\forall k \in {\cal K}$, but the cloud server receives a noisy version of $\frac{1}{K} \sum\nolimits_{k=1}^K {\bm g}^{[l]}_k$.

\textbf{3) IRS-free channel}: Edge nodes communicate directly with the cloud server without the aid of the IRS.

\textbf{4) SDR}: In this scheme, the SDR method \cite{SDR} is applied to solve ${\textsf P}_8$, which is introduced to validate the effectiveness of the proposed DC algorithm.

The test accuracy w.r.t. FL training round is shown in Fig. \ref{TA-TrainingRound}. From this figure, we immediately observe that without the aid of the IRS, the all-node selection scheme performs poorly. In contrast, with the aid of the IRS, the performance of the all-node selection scheme improves significantly, demonstrating the great potential of introducing IRS to FL systems. We can also observe from this figure that the weight-selection based framework achieves a better performance than its node-selection based counterpart, and both frameworks exhibit enhanced performance compared to the all-node selection scheme. Furthermore, we also observe that the proposed DC algorithm is superior to the SDR method in optimizing the IRS configuration, thereby leading to improved FL performance.

\begin{figure}
\centering
\includegraphics[width = 8.6cm]{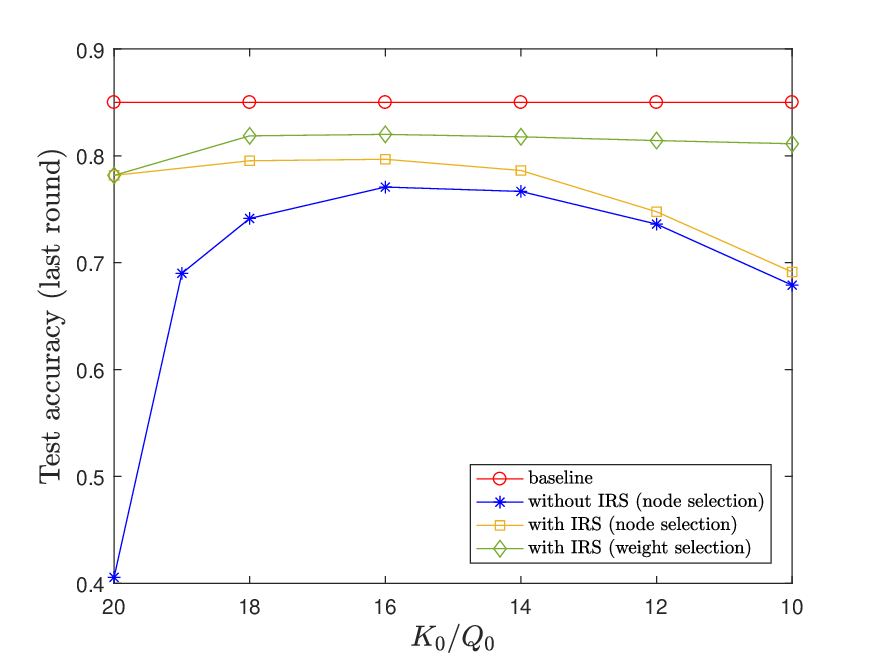}
\caption{Test accuracy w.r.t $K_0$ or equivalently $Q_0$.} \label{TA-DataSize}
\end{figure}

Fig. \ref{TA-DataSize} illustrates the test accuracy (after $L = 100$ training rounds) w.r.t. $K_0(Q_0)$. As depicted, an increase in $K_0(Q_0)$ from $10$ to $16$, leads to an improvement in the performance of both the node-selection and weight-selection based FL frameworks. However, when we further increase $K_0(Q_0)$ to $18$ and $20$, the performance of both frameworks begins to decline, demonstrating that there indeed exists a tradeoff between the model misfit loss and the communication error loss, as indicated by \eqref{Eq-NSFL-Convergence3} and \eqref{Eq-WSFL-Convergence3}. Moreover, we can also observe that the performance of the weight-selection based FL framework is robust to the increase of $Q_0$. In other words, we can randomly pick a value between $10$ and $18$ for $Q_0$ without significantly compromising the learning performance.

\begin{figure}
\vskip -4pt
\centering
\includegraphics[width = 8.6cm]{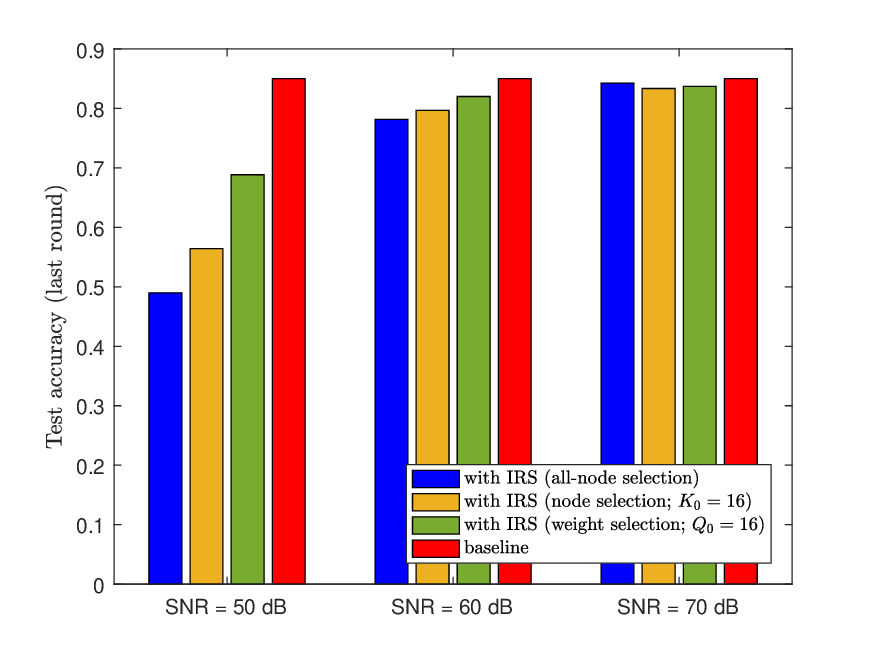}
\caption{Test accuracy w.r.t. $\textsf{SNR}$.} \label{TA-SNR}
\end{figure}

Fig. \ref{TA-SNR} shows the test accuracy (after $L = 100$ training rounds) w.r.t. $\textsf{SNR}$. It can be observed from this figure that when increasing $\textsf{SNR}$ from 50 dB to 70 dB, the learning performances of both node-selection and weight-selection based FL frameworks improve; the reason is that the communication errors decrease as $\textsf{SNR}$ increases. Moreover, the weight-selection based FL framework outperforms its node-selection based counterpart, especially when $\textsf{SNR}$ is low. We can also observe from this figure that when $\textsf{SNR} = 70$ dB, the performance of the all-node selection scheme improves significantly compared to that when $\textsf{SNR} = 50$ dB. The all-node selection scheme even outperforms the node-selection and weight-selection based frameworks when $\textsf{SNR} = 70$ dB, which means that we should select more edge nodes (i.e., larger $K_0$) or assign more edge nodes a large weight coefficient (i.e., larger $Q_0$) in model aggregation when $\textsf{SNR}$ is sufficiently high. That is the reason why the high and low $\textsf{SNR}$ regions are respectively termed data limited and $\textsf{SNR}$ limited regions in literature \cite{GuangxuZhu-TWC}.

\begin{figure}
\centering
\includegraphics[width = 8.6cm]{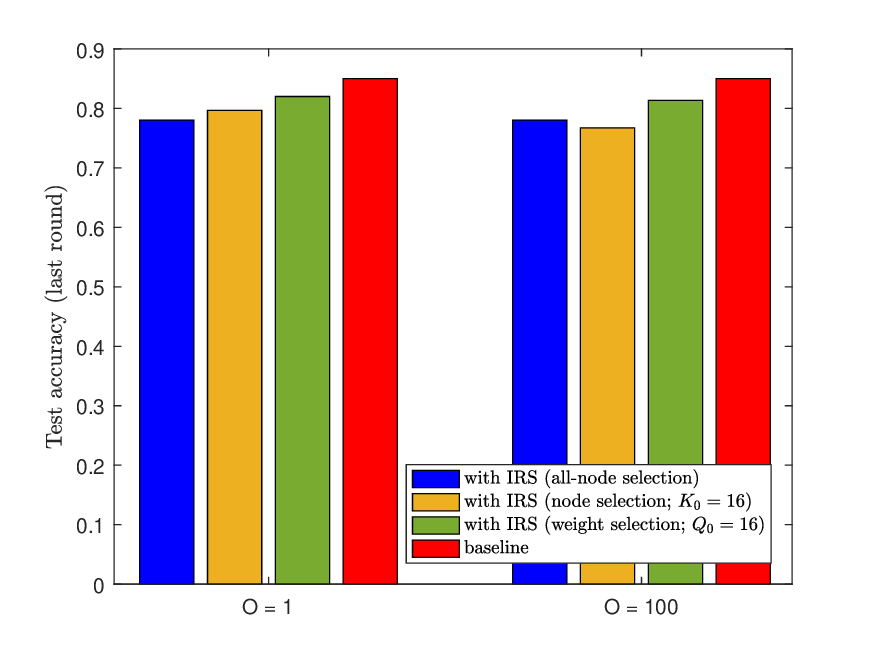}
\caption{Test accuracy w.r.t. channel coherence time.} \label{TA-Time}
\end{figure}

Fig. \ref{TA-Time} shows test accuracy w.r.t. coherence time of wireless links, which is evaluated by the number of FL training rounds, i.e., $O$. It can be observed from this figure that when the wireless links change slowly, i.e., $O = 100$, the node-selection based FL framework performs worse than its all-node selection based counterpart, though the former outperforms the latter when $O = 1$. The reason can be explained as follows. When $\{h_{d,k,j}\}$, $\{\bm h_{r,k,j}\}$, and $\{\bm z_j\}$ remain constant during $O = 100$ training rounds, the selected edge nodes in each FL round remain the same. In other words, some edge nodes are never involved in the FL learning process, and thus their unique data samples are discarded forever, which inevitably degrades the learning performance. This explains why complicated edge node scheduling algorithms are often needed in node-selection based FL systems \cite{JichaoLeng-WCL, HaoYang-TC}. In contrast, since all edge nodes are allowed to participate in model aggregation, the weight-selection based FL framework performs similarly in both $O = 1$ and $O = 100$.

\begin{figure}
\centering
\includegraphics[width = 8.6cm]{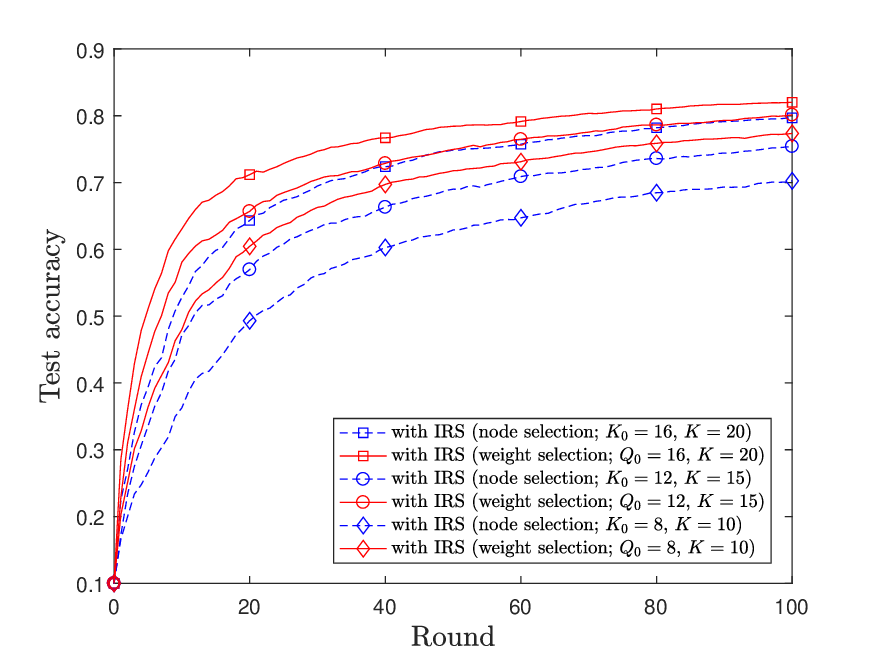}
\caption{Test accuracy w.r.t. number of edge nodes.} \label{TA-EdgeNodes}
\end{figure}

Furthermore, as the number of edge nodes increases, clear enhancement in the performance of both the node-selection based and weight-selection based frameworks is observed, as depicted in Fig. \ref{TA-EdgeNodes}. However, it is worth noting that the weight-selection based framework consistently outperforms its node-selection based counterpart. Specifically, when the total number of edge nodes is set to $10$, $15$, and $20$, and $80\%$ of them are selected in the node-selection based framework, the weight-selection based framework exhibits performance gains of $10.12\%$, $6.26\%$, and $2.35\%$, respectively.

\section{Conclusions} \label{Section-Conclusions}
In this paper, we have studied the AirComp-empowered model aggregation approach for IRS-assisted FL systems. We first considered the conventional node-selection based FL framework, analyzed its convergence, derived an upper bound on its performance, and introduced how to optimize the selected edge nodes, along with their transmit equalization coefficients, the IRS phase shifts, and the receive factors of the cloud server to minimize the MSE. We further proposed a weight-selection based FL framework to avoid the noticeable decrease in the learning performance caused by node selection. In such a framework, we assigned each edge node an optimized weight coefficient in model aggregation instead of discarding any of them. The theoretical analyses and numerical experiments revealed that the weight-selection based FL framework was superior to its node-selection based counterpart. Moreover, though both frameworks had a tradeoff between the achieved MSE and the fraction of data involved in learning, the weight-selection based framework was more robust to changes in the fraction of data.

\begin{appendices}
\section{}\label{Theorem1-Proof}
By substituting \eqref{Eq-ErrorInEq} and \eqref{Eq-NSEB} into \eqref{Eq-OneIter}, we have
\begin{align}\label{Eq-OneIter2}
    F(\bm w^{[l+1]}) & \le F(\bm w^{[l]}) - \frac{\left\|\nabla F(\bm w^{[l]})\right\|^2}{2\rho} + \frac{\big\|\bm e_1^{[l]}\big\|^2 + \big\|\bm e_2^{[l]}\big\|^2}{\rho} \nonumber \\[2ex]
    & \le F(\bm w^{[l]}) - \frac{\left\|\nabla F(\bm w^{[l]})\right\|^2}{2\rho} + \frac{\big\|\bm e_2^{[l]}\big\|^2}{\rho} \\[2ex]
    & + \frac{4}{\rho} \left(1 - \frac{\left|{\cal S}^{[l]}\right|}{K}\right)^2 \left(\gamma_1 + \gamma_2 \left\|\nabla F(\bm w^{[l]})\right\|^2\right). \nonumber
\end{align}
Next, based on \eqref{Eq-ConvexAssump} and \eqref{Eq-SmoothAssump}, we respectively derive that
\begin{align}
    \left\|\nabla F(\bm w^{[l]})\right\|^2 & \ge 2 \mu \left[F(\bm w^{[l]}) - F(\bm w^{\star})\right], \label{Eq-Grad-LB} \\[2ex]
    \left\|\nabla F(\bm w^{[l]})\right\|^2 & \le 2 \rho \left[F(\bm w^{[l]}) - F(\bm w^{\star})\right]. \label{Eq-Grad-UB}
\end{align}
By substituting \eqref{Eq-Grad-LB} and \eqref{Eq-Grad-UB} into \eqref{Eq-OneIter2}, we obtain that
\begin{align}\label{Eq-OneIter3}
    F(\bm w^{[l+1]}) \le F(\bm w^{[l]}) + \frac{4 \gamma_1}{\rho} \left(1 - \frac{\left|{\cal S}^{[l]}\right|}{K}\right)^2 + \frac{\big\|\bm e_2^{[l]}\big\|^2}{\rho} \nonumber \\[2ex]
    + \left[8 \gamma_2 \left(1 - \frac{\left|{\cal S}^{[l]}\right|}{K}\right)^2 - \frac{\mu}{\rho}\right] \left[F(\bm w^{[l]}) - F(\bm w^{\star})\right].
\end{align}
By first subtracting $F(\bm w^{\star})$ and then taking expectation on both sides of \eqref{Eq-OneIter3}, we can obtain
\begin{align}\label{Eq-OneIter4}
    \hspace{-0.25cm} & {\mathbb E}\left[F(\bm w^{[l+1]}) - F(\bm w^{\star})\right] \nonumber \\[2ex]
    \hspace{-0.25cm} & \le \frac{4 \gamma_1}{\rho} \left(1 - \frac{\left|{\cal S}^{[l]}\right|}{K}\right)^2 ~\overset{(a)}{+}~ \frac{1}{\rho} \frac{d}{\textsf{SNR} \left|{\cal S}^{[l]}\right|^2} \max\limits_{j \in {\cal N}} \left|m^{[l]}_j\right|^2 \nonumber \\[2ex]
    \hspace{-0.25cm} & + \left[8 \gamma_2 \left(1 - \frac{\left|{\cal S}^{[l]}\right|}{K}\right)^2 + 1 - \frac{\mu}{\rho}\right] {\mathbb E}\left[F(\bm w^{[l]}) - F(\bm w^{\star})\right],
\end{align}
where $(a)$ follows due to \eqref{Eq-CNEB}. Last, applying \eqref{Eq-OneIter4} recursively for $l = L, \cdots, 1$, we obtain \eqref{Eq-NSFL-Convergence} and complete the proof.

\section{}\label{Theorem-NonConvex-Proof}
By setting $|{\cal S}^{[l]}| = K_0$, we reduce \eqref{Eq-OneIter2} to
\begin{align}\label{Eq-NonConvex-OneIter1}
    F(\bm w^{[l+1]}) & \le F(\bm w^{[l]}) - \frac{a_0}{2\rho} \left\|\nabla F(\bm w^{[l]})\right\|^2 \nonumber \\[2ex]
    & + \frac{4 \gamma_1}{\rho} \left(1 - \frac{K_0}{K}\right)^2 + \frac{\big\|\bm e_2^{[l]}\big\|^2}{\rho}.
\end{align}
Summing both sides of \eqref{Eq-NonConvex-OneIter1} for $\forall l = 1, \cdots, L$, we obtain that
\begin{align}\label{Eq-NonConvex-OneIter2}
    F(\bm w^{\star}) & \le F(\bm w^{[L+1]}) \le F(\bm w^{[1]}) - \frac{a_0}{2\rho} \sum\limits_{l = 1}^L \left\|\nabla F(\bm w^{[l]})\right\|^2 \nonumber \\[2ex]
    & + \frac{4 \gamma_1 L}{\rho} \left(1 - \frac{K_0}{K}\right)^2 + \frac{1}{\rho} \sum\limits_{l = 1}^L \big\|\bm e_2^{[l]}\big\|^2.
\end{align}
After slight manipulations, we have
\begin{align}\label{Eq-NonConvex-OneIter3}
     \frac{1}{L} \sum\limits_{l = 1}^L \left\|\nabla F(\bm w^{[l]})\right\|^2 & \le \frac{2 \rho}{a_0 L} [F(\bm w^{[1]}) - F(\bm w^{\star})] \\[2ex]
     & + \frac{8 \gamma_1}{a_0} \left(1 - \frac{K_0}{K}\right)^2 + \frac{2}{a_0 L} \sum\limits_{l = 1}^L \big\|\bm e_2^{[l]}\big\|^2. \nonumber
\end{align}
Taking expectation on both sides of \eqref{Eq-NonConvex-OneIter3}, we can obtain \eqref{Eq-NonConvex-Convergence} and complete the proof.

\section{}\label{Proposition1-Proof}
Once $\bm \theta$ and $\cal S$ are given, we can simplify ${\textsf P}_1$ as
\begin{subequations}\label{OP2}
    \begin{eqnarray}
        & \min\limits_{x, \{x_j\}} & x \\
        & \text{s.t.} & 0< x_j \le x, ~\forall j \in {\cal N}, \\[1ex]
        & & \frac{1}{N} \sum\limits_{j=1}^N \frac{\beta_{k,j}}{x_j \left|h_{e, k,j} \right|^2} \le 1, ~\forall k \in {\cal S},
\end{eqnarray}
\end{subequations}
where we have defined $x_j = |m_j|^2$, $\forall j \in {\cal N}$, and $x = \max\{x_j\}$. Then, we write the Lagrange of \eqref{OP2} as
\begin{align}\label{OP2-Lagrange}
    {\cal L} & = x + \sum\limits_{j=1}^N \mu_j (x_j - x) \nonumber \\[1ex]
    & + \sum\limits_{k \in \cal S} \nu_k \left(\frac{1}{N} \sum\limits_{j=1}^N \frac{\beta_{k,j}}{x_j \left|h_{e, k,j} \right|^2} - 1\right),
\end{align}
where $\mu_j \ge 0$, and $\nu_k \ge 0$, are the associated Lagrange multipliers. Then, the Karush-Kuhn-Tucker (KKT) condition for \eqref{OP2-Lagrange} are given by
\begin{subequations}\label{OP2-KKT}
\begin{eqnarray}
    \frac{\partial {\cal L}}{\partial x} = 1 - \sum\limits_{j=1}^N \mu_j = 0; \label{OP2-KKTa} \\
    \frac{\partial {\cal L}}{\partial x_j} = \mu_j - \sum\limits_{k \in \cal S} \frac{\nu_k}{N} \frac{\beta_{k,j}}{x^2_j \left|h_{e, k,j} \right|^2} = 0, ~\forall j \in {\cal N}; \label{OP2-KKTb}\\[1ex]
    \mu_j (x_j - x) = 0, ~\forall j \in {\cal N};\label{OP2-KKTd}\\[1ex]
    \mu_j \ge 0, ~\forall j \in {\cal N}; \label{OP2-KKTe} \\[1ex]
    \nu_k \ge 0, ~\forall k \in {\cal S}. ~\label{OP2-KKTf}
\end{eqnarray}
\end{subequations}
Suppose that there exists an $x_l < x$, $\exists l \in {\cal N}$. According to \eqref{OP2-KKTd}, we have $\mu_l = 0$. As a result, we obtain
\begin{equation}
    \sum\limits_{k \in \cal S} \frac{\nu_k}{N} \frac{\beta_{k,l}}{x^2_l \left|h_{e, k,l} \right|^2} = \mu_l = 0,
\end{equation}
which implies that $\nu_k = 0$, $\forall k \in \cal S$.

On the other hand, according to \eqref{OP2-KKTa}, at least one of $\mu_1, \cdots, \mu_N$ should be larger than zero, e.g., $\mu_p > 0$, where $p \in {\cal N}\slash\{l\}$. Recalling \eqref{OP2-KKTb}, we have
\begin{equation}\label{OP2-Contradiction}
    \sum\limits_{k \in \cal S} \frac{\nu_k}{N} \frac{\beta_{k,p}}{x^2_p \left|h_{e, k,p} \right|^2} = \mu_p > 0.
\end{equation}
It can be observed from \eqref{OP2-Contradiction} that at least one $\nu_k$ should be larger than zero, which contradicts with the previous conclusion that $\nu_k = 0$, $\forall k \in \cal S$. Thus, we have proven that $x_l < x$ cannot exist, i.e., $x_1 = \cdots = x_N = x$. Moreover, recall the power constraint that
\begin{equation}
    \frac{1}{N} \sum\limits_{j=1}^N \frac{\beta_{k,j}}{x_j \left|h_{e,k,j} \right|^2} = \frac{1}{N} \sum\limits_{j=1}^N \frac{\beta_{k,j}}{x \left|h_{e,k,j} \right|^2} \le 1, ~\forall k \in {\cal S}.
\end{equation}
Consequently, we have
\begin{equation}
    x \ge \frac{1}{N} \sum\limits_{j=1}^N \frac{\beta_{k,j}}{\left|h_{e,k,j} \right|^2}, ~\forall k \in {\cal S},
\end{equation}
which completes the proof.
\end{appendices}

\ifCLASSOPTIONcaptionsoff
  \newpage
\fi

\end{document}